\documentclass[journal]{IEEEtran}
\usepackage{stfloats}
\usepackage{float}
\usepackage{graphicx}
\usepackage{subfigure}
\usepackage{color}
\usepackage{algpseudocode}
\usepackage{amssymb}
\usepackage{amsmath}
\usepackage{algorithm}
\usepackage{bm}
\usepackage[colorlinks]{hyperref}

\begin{document}

\title{A Framework of  FAS-RIS Systems: Performance Analysis and Throughput Optimization}

\author{Junteng Yao, Xiazhi Lai, Kangda Zhi, Tuo Wu, Ming Jin, Cunhua Pan,  Maged Elkashlan, \\Chau Yuen, \emph{Fellow, IEEE}, and Kai-Kit Wong, \emph{Fellow}, \emph{IEEE}

\vspace{-5mm}

\thanks{J. Yao and M. Jin are with the Faculty of Electrical Engineering and Computer Science, Ningbo University, Ningbo 315211, China (E-mail: $\rm \{yaojunteng, jinming\}@nbu.edu.cn$).  X. Lai is with the School of Computer Science, Guangdong University of Education, Guangzhou, Guangdong, China (E-mail: $\rm xzlai@outlook.com$).  K. Zhi is with the School of Electrical Engineering and Computer Science, Technical University of Berlin, 10623 Berlin (E-mail: $\rm k.zhi@tu\text{-}berlin.de$). T. Wu and C. Yuen are with the School of Electrical and Electronic Engineering, Nanyang Technological University, 639798, Singapore (E-mail: $\rm \{tuo.wu, chau.yuen\}@ntu.edu.sg$). C. Pan is with the National Mobile Communications Research Laboratory, Southeast University, Nanjing 210096, China (E-mail: $\rm cpan@seu.edu.cn$). M. Elkashlan is with the School of Electronic Engineering and Computer Science at Queen Mary University of London, London E1 4NS, U.K. (E-mail: $\rm maged.elkashlan@qmul.ac.uk$).  K.-K. Wong is with the Department of Electronic and Electrical Engineering, University College London, WC1E 6BT London, U.K., and also with the Yonsei Frontier Laboratory and the School of Integrated Technology, Yonsei University, Seoul 03722, South Korea (E-mail: $\rm kai\text{-}kit.wong@ucl.ac.uk$).}
}

\maketitle

\begin{abstract}
In this paper, we investigate reconfigurable intelligent surface (RIS)-assisted communication systems which involve a fixed-antenna base station (BS) and a mobile user (MU) that is equipped with fluid antenna system (FAS). Specifically, the RIS is utilized to enable communication for the user whose direct link from the base station is blocked by obstacles.  We propose a comprehensive framework that provides transmission design for both static scenarios with the knowledge of channel state information (CSI) and harsh environments where CSI is hard to acquire. It leads to two approaches: a CSI-based scheme where CSI is available, and a CSI-free scheme when CSI is inaccessible. Given the complex spatial correlations in FAS, we employ block-diagonal matrix approximation and independent antenna equivalent models to simplify the derivation of outage probabilities in both cases.  Based on the derived outage probabilities, we then optimize the throughput of the FAS-RIS system. For the CSI-based scheme, we first propose a gradient ascent-based algorithm to obtain a near-optimal solution. Then, to address the possible high computational complexity in the gradient algorithm, we approximate the objective function and confirm a unique optimal solution accessible through a bisection search method. For the CSI-free scheme, we apply the partial gradient ascent algorithm, reducing complexity further than full gradient algorithms. We also approximate the objective function and derive a locally optimal closed-form solution to maximize throughput.  Simulation results validate the effectiveness of the proposed framework for the transmission design in FAS-RIS systems.
\end{abstract}


\section{Introduction}
Multiple-input multiple-output (MIMO) technique played a crucial role in the development of wireless communications. The diversity gains provided by MIMO  heavily depend on the spatial positioning of antennas at intervals of at least half a wavelength \cite{CXWang23, BCetiner04}. However, devices with limited physical space, such as mobile phones and sensors in the Internet of Things (IoT), are often constrained to have only a few antennas, which significantly impairs communication performance. To overcome this limitation,  fluid antenna system (FAS) has recently emerged as a promising solution. Specifically, FAS utilizes a liquid-based antenna \cite{Shen-tap_submit2024} or reconfigurable pixel-based antenna \cite{7762757,9491941} that can switch freely to the optimal position of maximum signal strength within a prescribed space. Thanks to the flexibility of such reconfigurable antennas, FAS effectively exploits the spatial degrees of freedom (DoFs) to gain sufficient diversity and enhance communication performance significantly even in a tiny space \cite{XLai24,KKWong21,KKWong20,JYao24,MKhammassi23,WKNew242,LZhu24}.

Inspired by the promising feature of FAS, researchers have extensively integrated it into various wireless communication systems, including secure communication systems \cite{JD24,BTang23}, multiple access systems \cite{JZheng24,WKNew24,KKWong22}, and MIMO systems \cite{YYe23,WKNew231}. Specifically, the authors in \cite{JD24}  investigated secrecy outage probability in point-to-point communication for Nakagami-$m$ fading. Moreover, Tang \emph{et al.} maximized the secrecy rate by designing power allocation respectively under perfect and imperfect channel state information (CSI) scenarios \cite{BTang23}. Further advancing the field, Wong \emph{et al.} proposed the concept of fluid antenna multiple access (FAMA) for handling interference without CSI at the transmitter side. More recently,  Zheng {\em et al.} \cite{JZheng24} contributed by analyzing the average block error rate (BLER) in non-orthogonal multiple access (NOMA) systems with FAS, emphasizing short-packet communications in modern networks. Besides, New \emph{et al.} \cite{WKNew24} maximized the sum rate of NOMA systems by selecting optimal port and power allocation. Complementing these efforts,  the authors in \cite{YYe23,WKNew231} studied the FAS with multiple RF chains, which can significantly improve the performance  compared with conventional MIMO systems.

Although the aforementioned studies have validated the benefits of FAS, they have not addressed scenarios involving severe channel blocking due to obstacles or long distances between transmitters and receivers in hostile radio environments. Therefore, it is important to understand the detrimental impact of unfavorable propagation environment and wireless channel attenuation on the receivers in the FAS-assisted wireless communication systems. Fortunately, reconfigurable intelligent surface (RIS) has recently been proposed to address these kinds of issues, owing to its capability to customize the propagation environment and bypass the obstacles \cite{Ghadi-2024wcl,WuQ1,KZhi22}. RIS can adjust its reflecting elements to reflect impinging signals in a constructive and cost-efficient way, thereby introducing additional diversity to enhance communication system performance. Additionally, RIS can be flexibly deployed on the facade of high buildings, which can enable high-quality line-of-sight (LoS) communication to users whose direct links from the base station (BS) are blocked. This ability to ``reconfigure'' the radio propagation environment has recently garnered significant research interest in RIS-assisted mobile edge computing (MEC) systems \cite{TBai20,YYang22}, RIS-assisted wireless power transfer (WPT) systems \cite{WuQ3,CPan20}, RIS-assisted NOMA systems \cite{YLi21,JZhu21}, and RIS-assisted proactive monitoring systems \cite{JYao20,GHu23}.

Encouraged by the capability of FAS and RIS for enhancing wireless communication performance from different perspectives, it remains a compelling question to analyze and unveil the theoretical performance of integrated FAS-RIS systems \cite{Ghadi-2024wcl}. Building upon the theoretical analysis, it is also meaningful to conduct optimization design for the throughput performance in FAS-RIS systems, which is also an unknown task.

Besides, it is worthy noting that the works on the performance analysis of either RIS or FAS are mostly based on the assumption of perfect knowledge of channel information. However, both the passive nature of the RIS and the characteristic of frequently moving position of the fluid antenna would lead to the high difficulty in channel acquisition, requiring large number of pilots and resulting in huge overhead.  In the static scenario, the channel coherence time could be long which may support high-accuracy channel acquisition for FAS-RIS systems and support some CSI-based algorithms to achieve high performance. However, in scenarios with mobility, the channel coherence time could be short which may not afford the heavy channel estimation overhead in FAS-RIS systems. As a result, perfect CSI-based algorithms cannot work and a robust algorithm capable of tackling this harsh condition is highly necessary. Therefore, when studying FAS-RIS systems, a comprehensive framework providing transmission design solutions for both CSI-based ideal scenarios and CSI-free harsh scenarios is crucial.

Nevertheless, investigating integrated FAS-RIS systems is a non-trivial task due to several inherent challenges. On one hand,  the characteristic of the spatial correlation between the ports of FAS significantly complicates the derivation of outage probability, leading to the difficulty of performance analysis. Given the intricate expression of the outage probability, the subsequent optimization design with respect to throughput is a challenging problem that cannot be addressed using conventional optimization methods. On the other hand, the CSI-free case further exacerbates the outage probabilities, making them even more complex than the CSI-based case. Accordingly, with a more intricate expression of outage probability, the task of optimizing throughput becomes substantially more difficult.

In light of the aforesaid background, in this paper, we propose a comprehensive framework that includes performance analysis and throughput optimization for FAS-RIS systems. The framework is designed to accommodate both scenarios: with and without CSI.  It comprises two distinct schemes: a CSI-based scheme, and a CSI-free scheme. To facilitate the derivation of outage probabilities, we introduce two models: the block-diagonal matrix approximation model and the independent antenna equivalent model, which enable us to derive closed-form approximations of outage probabilities. {Additionally, given these refined expressions of outage probabilities, we utilize the gradient ascent, bisection search, and closed-form solution algorithms to optimize the throughput.} The main contributions of this paper are summarized as follows:

\begin{itemize}
\item \textbf{\emph{Novel System Model}}: We study a FAS-RIS system where the BS communicates to a fluid antenna-equipped mobile user (MU) with the help of RIS. We consider two possible scenarios whether the BS has/does not have the  CSI of the BS-RIS-MU link. With CSI perfectly known, the BS can design the optimal reflecting pattern for RIS, while the reflecting elements of the RIS are set randomly to realize a low-overhead design when CSI is unknown. We also adopt the general Rician fading model to characterize the existence of both LoS and NLoS paths.

\item \textbf{\emph{CSI-Based Scheme}}: With access to CSI, we propose a CSI-based scheme that integrates performance analysis and throughput optimization. Utilizing realistic spatial correlation models, we employ both the block-diagonal matrix approximation model and the independent antenna equivalent model, coupled with the Lyapunov central limit theorem (CLT), to derive a closed-form approximation of the outage probability.  Building upon this, we apply the gradient ascent algorithm to achieve a near-optimal solution for throughput optimization. Subsequently, to reduce computational complexity, we approximate the objective function and provide a theoretical proof demonstrating the existence of a unique optimal solution for the formulated problem.

\item \textbf{\emph{CSI-free Scheme}}: To address the scenarios with high CSI-acquisition difficulties, we propose a low-overhead CSI-free scheme, analyze its performance, and optimize the throughput.  Initially, we employ the block-diagonal matrix approximation model and independent antenna equivalent model, coupled with the  CLT to derive  a closed-form approximation of the outage probability.  Then, we propose using the partial gradient ascent algorithm for throughput optimization, which employs the bisection search method within a partial interval of the optimization variable and utilizes the gradient ascent algorithm in the residual interval. To further reduce the computational complexity, we propose an approximate method that yields a closed-form solution.

\item  \textbf{\emph{Confirmation of effectiveness and robustness}}: Simulation results demonstrate the superior approximate accuracy of the proposed schemes. Besides, the proposed optimization algorithms showcase enhanced performance compared to benchmark methods, confirming the robustness of the proposed framework. Furthermore, it is shown that the proposed CSI-free scheme can achieve satisfactory performance. With limited channel coherence time, it can outperform CSI-based design after considering the practical channel estimation overhead.
\end{itemize}

The rest of this paper is organized as follows. Section \ref{sec:system} presents our model of the FAS-RIS system. In Section \ref{sec:csi}, the CSI-based scheme is proposed to obtain the approximate outage probability and optimize the throughout for scenarios where CSI is available. Section \ref{sec:beyond} proposes a CSI-free scheme to derive the the approximate outage probability and optimize the throughout for scenarios where CSI is not accessible. Finally, our numerical results are provided in Section \ref{sec:results}, and our conclusions are drawn in Section \ref{sec:conclude}.

\section{System Model}\label{sec:system}
\begin{figure}[h]
\centering
\includegraphics[width=3.5in]{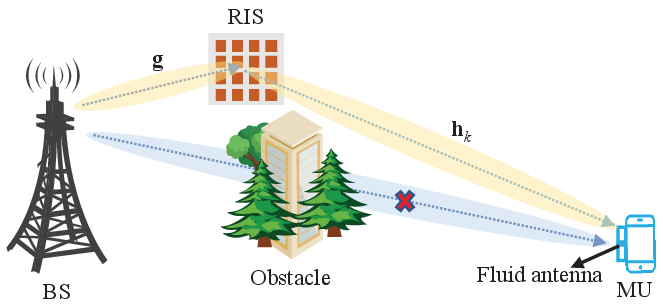}
\caption{The system model of the FAS-RIS communication system.}\label{fig:sm}
\end{figure}

As depicted in Fig.~\ref{fig:sm}, we consider a RIS-assisted downlink communication system comprising a  BS with a single fixed-position antenna, a RIS with $M$ reflecting elements, and a MU equipped with a fluid antenna  capable of switching among $N$ ports within a linear space of $W\lambda$, where $W$ is the normalized length and $\lambda$ is the wavelength. The direct link between the BS and MU is assumed broken by obstacles such as buildings, or natural barriers \cite{KZhi221,KZhi222}. To re-establish communication between the BS and MU, the RIS is utilized to enable signal transmission from the BS to the MU. Moreover, we assume that delays resulting from port switching are negligibly small, e.g., using reconfigurable pixel antennas \cite{7762757,9491941}.

\subsection{Communication Model}
As shown in Fig.~\ref{fig:sm}, the RIS is deployed close to the BS and positioned at high altitude \cite{KZhi221}. Therefore, the BS-RIS link is considered to LoS. Denoting the channel response vector of the BS-RIS link as $\mathbf{g}\in{\mathbb{C}^{1\times M}}$, we have
\begin{align}\label{eq1}
\mathbf{g}= \sqrt{\beta}\bar{\mathbf{g}},
\end{align}
where $\beta$ denotes the path-loss of the BS-RIS channel, and $\bar{\mathbf{g}}$ represents the LoS channel, which is modeled as the steering vector of a two-dimensional uniform squared planar array (USPA).

Given the fact that the RIS is closer to the BS and more distant from the MU, the RIS-MU link invariably includes NLoS components. Consequently, we adopt the Rician model for the channel between the RIS and MU. Denoting the channel  response vector connecting the RIS to the $k$-th port of the MU's fluid antenna as $\mathbf{h}_k\in{\mathbb{C}^{1\times{M}}}$, we have
\begin{align}\label{eq2}
\mathbf{h}_k = \sqrt{\frac{\alpha K}{K+1}}\bar{\mathbf{h}}_k + \sqrt{\frac{\alpha}{K+1}}\tilde{\mathbf{h}}_k,
\end{align}
in which $\alpha$ represents the path loss, $K$ denotes the Rician factor, $\bar{\mathbf{h}}_k$ denotes the LoS component, and $\tilde{\mathbf{h}}_k$ comprises the NLoS components, whose elements are independently and identically distributed (i.i.d.) zero-mean complex Gaussian random variables (RVs) with unit variance.

Let us further denote the reflection matrix of the RIS as $\mathbf{\Phi}=\mathrm{diag}\{\phi_1, \dots, \phi_M\}$, with each $\phi_m=\exp(j\theta_m)$ for $m\in\mathcal{M} =\{1,\dots,M\}$, where $\theta_m\in[0,2\pi)$ represents the phase shift of the $m$-th element. Subsequently, the  cascaded channel   of the  BS-RIS-MU link is expressed as
\begin{align}\label{eqqe2}
 {f}_k = \mathbf{h}_k\mathbf{\Phi}\mathbf{g}^H.
\end{align}

Accordingly, the received signal  at the $k$-th port within the FAS of the MU can be formulated as
\begin{align}\label{eq3}
y_k &=\sqrt{P}  {f}_kx+n \nonumber\\
 &=\sqrt{P}\mathbf{h}_k\mathbf{\Phi}\mathbf{g}^Hx+n, ~k\in\mathcal{N}=\{1,2,\dots,N\},
\end{align}
where $\mathbb{E}[|x|^2]=1$ and $P$ is the transmit power of the BS. $n\sim\mathcal{CN}(0,\sigma^2)$ is the additive noise at the MU.

\subsection{Outage Probability}
Building upon \eqref{eq3}, the signal-to-noise ratio (SNR) at the $k$-th port of FAS within the MU can be written as
\begin{align}\label{eq6}
\gamma_{k}&=\frac{PA^2_k}{\sigma^2}, ~\mbox{for }k\in\mathcal{N},
\end{align}
{where $A_k=|f_k|$.} Using FAS, the port exhibiting the highest channel gain can be selected for receiving the signal.  This leads to  the maximum of $A_k$, which is expressed as
\begin{align}\label{eq7}
A_{\max} =\max\{A_1, A_2, \dots, A_N\}.
\end{align}
For a comprehensive analysis of the proposed RIS-FAS communication systems, this paper considers the following two scenarios based on whether the BS can obtain the instantaneous CSI of the BS-RIS-MU link.
 \subsubsection{With CSI}
 In the first scenario, where the channel coherence time is long and enable BS to acquire the perfect CSI,  the phase shift matrix $\mathbf{\Phi}$ can be designed to optimize system performance  \cite{WuQ1,KZhi22,JYao20}. Consequently,  $A_k$   is calculated as
\begin{align}\label{eq10}
A_k= \sqrt{\beta}\sum_{i=1}^M|h_k^i|, ~\mbox{for }k\in\mathcal{N}.
\end{align}

\subsubsection{Without CSI}
In the second scenario, if the environment is relatively dynamic, there are limited time slots in each channel coherence time. To avoid the prohibitive overhead and guarantee enough time for data transmission, a low-overhead scheme is considered here, which will be proven to have satisfactory performance. Specifically, $\phi_m$  is assumed to be set randomly and follow a uniform distribution. As a result, $A_k$ is determined by
$A_k=| {f}_k|$.

\subsubsection{Outage Probability Formulation}
Subsequently, the SNR for the MU can be formulated as
\begin{align}\label{eq8}
\gamma&=\frac{P|A_{\max}|^2}{\sigma^2}.
\end{align}
Given a specific transmission rate $R$, the outage probability at the MU is formulated as
\begin{align}\label{eq9}
\mathbb{P}^{\mathrm{out}}&=\mathbb{P}(\log_2(1+\gamma)<R)=\mathbb{P}\left(A_{\max}<\sqrt{\gamma_{\rm th}}\right),
\end{align}
where $\gamma_{\rm th}=(2^R-1)\sigma^2/P$.

\subsection{FAS Channel Correlation Model}
Given the close proximity of the ports in FAS, the channels, $\mathbf{h}_k$, are inherently correlated. To allow the analysis of outage probability $\mathbb{P}^{\mathrm{out}}$, this paper uses the Jakes' model to characterize this correlation. The correlation coefficient between the first port and the $k$-th port is modeled as \cite{KKWong23}
\begin{align}\label{eq4}
\mu_{1,k} &=J_0\left(\frac{2\pi(k-1)}{N-1}W\right), k\in\mathcal{N},
\end{align}
where $J_0(\cdot)$ is the zero-order Bessel function of the first kind. Consequently, the correlation coefficient matrix of the random vector $\mathbf{h}_m=[{h}_m(1),\dots,{h}_m(N)]$ is a Toeplitz matrix \cite{PR24}, which is given by
\begin{align}\label{eq5}
\mathbf{\Sigma}\in \mathbb{R}^{N\times N} & =\mathbf{topelitz}(\mu_{1,1},\mu_{1,2},\dots, \mu_{1,N})\notag\\
&=\left(
         \begin{array}{cccc}
         \mu_{1,1} &\mu_{1,2} & \cdots & \mu_{1,N} \\
        \mu_{1,2} & \mu_{1,1} & \cdots & \mu_{1,N-1} \\
        \vdots &  & \ddots & \vdots \\
       \mu_{1,N}& \mu_{1,N-1} & \cdots &\mu_{1,1} \\
         \end{array}
       \right).
\end{align}

\subsection{Block-Correlation Matrix Approximation Model}
Deriving the outage probability directly from the Toeplitz matrix $\mathbf{\Sigma}$ presents a  challenge due to its complex structure. To facilitate a more tractable analysis, we adopt an innovative approach by approximating the spatial correlation structure of the $N$ ports in the FAS using a block-diagonal correlation matrix \cite{PR24}, which is given by
\begin{equation}\label{eq11}
\hat{\mathbf{\Sigma}}\in \mathbb{R}^{N\times N}=
   \left[ \begin{array}{cccc}
\mathbf{C}_1 & \mathbf{0} &\cdots  &  \mathbf{0}  \\
\mathbf{0} & \mathbf{C}_2 &\cdots  &  \mathbf{0}  \\
 \vdots &  \vdots &\ddots & \vdots \\
\mathbf{0} &\mathbf{0} & \cdots & \mathbf{C}_B\\
\end{array}\right],
\end{equation}
where each submatrix $\mathbf{C}_b$ represents a constant correlation matrix of size $L_b$ and correlation $\mu^2_b$, defined as
\begin{equation}\label{eq3a}
\mathbf{C}_b\in \mathbb{R}^{L_b\times L_b}=
   \left[ \begin{array}{cccc}
1& \mu^2_b &\cdots  &  \mu^2_b  \\
\mu^2_b & 1 &\cdots  &  \mu^2_b \\
 \vdots &  \vdots &\ddots & \vdots \\
\mu^2_b &\mu^2_b & \cdots &1 \\
\end{array}\right], b\in\mathcal{B},
\end{equation}
where $\mathcal{B}=\{1,\dots,B\}$, and $\sum_{b=1}^B L_b=N$. Here, $B=|S\{\lambda_{\rm th}\}|$ is determined by the number of significant eigenvalues of $\bf\Sigma$, and $S\{\lambda_{\rm{th}}\} = \{\lambda_n | \lambda_n \geq \lambda_{\rm{th}}, n=1, \dots, N\}$, where $\lambda_{\rm{th}}$ is set to a small value to ensure that a sufficient number of eigenvalues are included in $S\{\lambda_{\rm{th}}\}$. The size of each block, $L_b$, is determined based on the optimization criterion \cite{PR24}
\begin{align}\label{d3}
\arg\min_{L_1, \dots, L_B} \mathrm{dist}(\hat{\mathbf{\Sigma}}, \mathbf{\Sigma}),
\end{align}
where $\mathrm{dist}(\cdot)$ represents a distance metric between two matrices. This metric quantifies the divergence of their eigenvalues, and the detailed procedure can be found in \cite{PR24}.

By incorporating $\mathbf{C}_b$, for $L_b$ ports, the channels $\mathbf{h}_k$, for all $k$, are inherently correlated. Their mathematical expressions can be formulated as \cite{KKWong20}
\begin{align}\label{eq3b}
\mathbf{h}_k =\sqrt{\frac{\alpha K}{K+1}}\bar{\mathbf{h}}_k+ \mu_b\tilde{\mathbf{h}}_b+ \sqrt{1-\mu_b^2}\mathbf{e}_k , k \in\mathcal{N},
\end{align}
where $\bar{\mathbf{h}}_k = [\bar{{h}}^1_k,\dots,\bar{{h}}^M_k]$, and $\tilde{\mathbf{h}}_b = [\tilde{{h}}^1_b,\dots,\tilde{{h}}^M_b]$ and $\mathbf{e}_k = [{e}^1_k,\dots,{e}^M_k]$ are i.i.d.~RVs, each following a complex normal distribution $\mathcal{CN}\left(\mathbf{0},\frac{\alpha}{K+1}\mathbf{I}_M\right)$.

Considering the two scenarios: whether the RIS is configured based on CSI or not, the approximation of the outage probability can be distinctly categorized into two cases using the block-correlation matrix approximation. The specifics of each scenario  are detailed in the following sections.

\section{CSI-Based Scheme}\label{sec:csi}
In this section, it is assumed that the BS can accurately estimate the instantaneous CSI of the BS-RIS-MU link, e.g., in some static scenarios. Given this capability, the phase shift matrix $\mathbf{\Phi}$ can be optimized accordingly \cite{WuQ1,KZhi22,JYao20}. Building on this assumption, we propose a CSI-based scheme to analyze the theoretical performance and optimize the throughput related to this configuration.

\subsection{Performance Analysis}
For the $b$-th submatrix of $\hat{\bm{\Sigma}}$, conditioned on $\tilde{h}_b^i$ for $\{i \in \mathcal{M} = \{1, \dots, M\}\}$, the distribution of $h_k^i$ is given by
\begin{align}\label{Aq1}
h_k^i \sim \mathcal{CN}\left(\mu_b \tilde{h}_b^i + \sqrt{\frac{\alpha K}{K+1}} \bar{h}_k^i, \frac{(1-\mu_b^2)\alpha}{K+1}\right), k \in \mathcal{N},
\end{align}
where $\mu_b$ is the correlation coefficient, and $\alpha$ and $K$ are scaling parameters. Under this conditioning,  $|h_k^1|, |h_k^2|, \dots, |h_k^M|$ follow independent Rician distributions.

The number of reflecting elements of the RIS are supposed to be sufficiently large to compensate the product pathloss attenuation and improve the communication performance. Subsequently, Lyapunov CLT \cite[Th. 6.2]{BKnaeble} can be employed so that the approximate distribution for $A_k$ can be derived as
\begin{align}\label{Aq2}
A_k \sim \mathcal{N}(\mu_k, \sigma_k^2), ~k \in \mathcal{N},
\end{align}
with the mean and variance expressed, respectively, by
\begin{align}
\mu_k &= \sqrt{\beta} \sum_{i=1}^M \frac{\sqrt{\pi \sigma_{0}^2}}{2} L_{\frac{1}{2}}\left(-\frac{v_{k,i}^2}{\sigma_{0}^2}\right),\label{Aq3}\\
\sigma_k^2 &= \beta \left(\sum_{i=1}^M \sigma_{0}^2 + v_{k,i}^2 - \frac{\pi \sigma_{0}^2}{4} L_{\frac{1}{2}}^2\left(-\frac{v_{k,i}^2}{\sigma_{0}^2}\right)\right),\label{Aq4}
\end{align}
where $v_{k,i} = \left|\mu_b \tilde{h}_b^i + \sqrt{\frac{\alpha K}{K+1}} \bar{h}_k^i\right|$, $\sigma_{0}^2 = \frac{(1-\mu_b^2)\alpha}{K+1}$, and $L_{\frac{1}{2}}(\cdot)$ represents the Laguerre polynomial.

Conditioned on $\mathbf{h}_b$, the joint probability density function (PDF) of $A_1, A_2, \dots, A_N$ is given as
\begin{align}\label{Aq5}
f_{A_1, \dots, A_N | \mathbf{h}_b}(r_1, \dots, r_N| \mathbf{r}_b)=&\prod_{k\in \mathcal{K}_b}\frac{1}{\sqrt{2\pi \sigma^2_k}} e^{-\frac{(r_k-\mu_k)^2}{2\sigma^2_k}}.
\end{align}
Thus, the joint PDF of $\mathbf{h}_b, A_1, A_2, \dots, A_N$ is expressed as
\begin{align}\label{Aq6}
&f_{\mathbf{h}_b, A_1, A_2, \dots, A_N}(\mathbf{r}_b, \dots, r_N)\nonumber\\
&=\frac{e^{-(\mathbf{r}_b-\bm{\mu}_b)^H\mathbf{R}^{-1}_b(\mathbf{r}_b-\bm{\mu}_b)}}{\pi^M|\mathbf{R}_b|}\prod_{k\in \mathcal{K}_b}\frac{1}{\sqrt{2\pi \sigma^2_k}} e^{-\frac{(r_k-\mu_k)^2}{2\sigma^2_k}},
\end{align}
where $\bm{\mu}_b=\mathbf{0}$, $\mathbf{R}_b=\sigma^2_b\mathbf{I}_M$, and $\sigma^2_b = \frac{\alpha}{K+1}$.
\begin{figure*}[hb]
  \centering
  \hrulefill
\begin{align}\label{Aq7}
&F_{\mathbf{h}_b, A_1, A_2, \dots, A_N}(\mathbf{r}_b, r_1, \dots, r_N)\nonumber\\
&=\int_{-\infty}^{\infty} \frac{e^{-(\mathbf{r}_b-\bm{\mu}_b)^H\mathbf{R}^{-1}_b(\mathbf{r}_b-\bm{\mu}_b)}}{\pi^M|\mathbf{R}_b|}\underbrace{\int_{0}^{\gamma_1} \cdot\cdot\cdot \int_{0}^{\gamma_N}}_{r_1, \dots, r_N}\prod_{k\in \mathcal{K}_b}\frac{1}{\sqrt{2\pi \sigma^2_k}} e^{-\frac{(r_k-\mu_k)^2}{2\sigma^2_k}} dr_1, \dots, dr_N, d\mathbf{r}_b\nonumber\\
&=\int_{-\infty}^{\infty} \frac{e^{-\frac{(\mathbf{r}_b-\bm{\mu}_b)^H(\mathbf{r}_b-\bm{\mu}_b)}{\sigma_b^2}}}{(\sigma^2_b\pi)^M}\prod_{k\in \mathcal{K}_b}\int_{0}^{\gamma_k} \frac{1}{\sqrt{2\pi \sigma^2_k}} e^{-\frac{(r_k-\mu_k)^2}{2\sigma^2_k}}dr_k, d\mathbf{r}_b\nonumber\\
&=\int_{-\infty}^{\infty} \frac{e^{-\frac{(\mathbf{r}_b-\bm{\mu}_b)^H(\mathbf{r}_b-\bm{\mu}_b)}{\sigma_b^2}}}{(\sigma^2_b\pi)^M}\prod_{k\in \mathcal{K}_b}\left(\int_{-\infty}^{\gamma_k} \frac{1}{\sqrt{2\pi \sigma^2_k}} e^{-\frac{(r_k-\mu_k)^2}{2\sigma^2_k}}dr_k-\int_{-\infty}^{0} \frac{1}{\sqrt{2\pi \sigma^2_k}} e^{-\frac{(r_k-\mu_k)^2}{2\sigma^2_k}}dr_k\right)d\mathbf{r}_b\nonumber\\
&=\int_{-\infty}^{\infty} \frac{e^{-\frac{(\mathbf{r}_b-\bm{\mu}_b)^H(\mathbf{r}_b-\bm{\mu}_b)}{\sigma_b^2}}}{(\sigma^2_b\pi)^M}\prod_{k\in \mathcal{K}_b}\left(\frac{1}{2}\mathrm{erf}\left(\frac{\gamma_k-\mu_k}{\sqrt{2\sigma^2_k}}\right)-\frac{1}{2}\mathrm{erf}\left(-\frac{\mu_k}{\sqrt{2\sigma^2_k}}\right)\right)d\mathbf{r}_b
\end{align}
\end{figure*}

Building upon \eqref{Aq6}, the joint cumulative distribution function (CDF) of $\mathbf{h}_b, A_1, A_2, \dots, A_N$ is derived in \eqref{Aq7} (see bottom of this page). Then, by substituting $\gamma_1=\cdots=\gamma_N=\sqrt{\gamma_{\rm th}}$ into the joint CDF in \eqref{Aq7}, the outage probability can be approximated as $\hat{\mathbb{P}}^{\mathrm{out}}$, which is formulated as
\begin{align}\label{Aq8}
\hat{\mathbb{P}}^{\mathrm{out}}=&\mathbb{P}(\log_2(1+\gamma)<R)\nonumber\\
=&\mathbb{P}\left(|A_{\max}|<\sqrt{\gamma_{\rm th}}\right)\nonumber\\
=&\prod_{b=1}^B\int_{-\infty}^{\infty} \frac{e^{-\frac{\mathbf{r}_b^H\mathbf{r}_b}{\sigma_b^2}}}{(\sigma^2_b\pi)^M}\times\prod_{k\in \mathcal{K}_b}\nonumber\\
&\left(\frac{1}{2}\mathrm{erf}\left(\frac{\sqrt{\gamma_{\rm th}}-\mu_k}{\sqrt{2\sigma^2_k}}\right)-\frac{1}{2}\mathrm{erf}\left(-\frac{\mu_k}{\sqrt{2\sigma^2_k}}\right)\right)d\mathbf{r}_b,
\end{align}
where $\mathrm{erf}(\cdot)$ is the error function.

The outage probability expression in \eqref{Aq8} involve integrals, resulting in complex computations. To simplify the  expression and facilitate analysis, we propose the independent antenna equivalent model, which eliminates the need of integrals in the approximations. The core concept of this model is to assume perfect correlation by setting $\mu_b \rightarrow 1$. Under this assumption, the block-correlation channels are treated as $B$ i.i.d.~channels, denoted by $\mathbf{h}_b$ for each $b \in \mathcal{B}$, with their detailed expressions provided in \eqref{eq2}.

\emph{Remark 1}: The outage probability derived from the independent antenna equivalent model serve as the upper bound for those obtained through the block-correlation matrix approximation model, provided that the values of $L_b$ for each $b \in \mathcal{B}$ are identical in both models. This finding aligns with the results provided in \cite{PR24}.

Let us use Lyapunov CLT to approximate $A_b, b\in\mathcal{B}$, which is given by
\begin{align}\label{Bq1}
A_b\sim \mathcal{N}(\bar{\mu},\bar{\sigma}^2),
\end{align}
with
\begin{align}
\bar{\mu}&=\sqrt{\beta}\sum_{i=1}^M\frac{\sqrt{\pi\sigma_{b}^2}}{2}L_{\frac{1}{2}}\left(-\frac{v_{i}^2}{\sigma_{b}^2}\right),\label{Bq2}\\
\bar{\sigma}^2&=\beta\left(\sum_{i=1}^M\sigma_{b}^2+v_{i}^2-\frac{\pi\sigma_{b}^2}{4}L^2_{\frac{1}{2}}\left(-\frac{v_{i}^2}{\sigma_{b}^2}\right)\right),\label{Bq3}
\end{align}
where $v_{i}=\left|\sqrt{\frac{\alpha K}{K+1}}\bar{h}_k^i\right|$.

Because $A_1, A_2, \dots, A_B$ are mutually independent, the joint PDF of $A_1, A_2, \dots, A_B$ is formulated as
\begin{align}\label{Bq4}
f_{A_1, \dots, A_B}(r_1, \dots, r_B)=&\prod_{b=1}^B\frac{1}{\sqrt{2\pi \bar{\sigma}^2}} e^{-\frac{(r_b-\bar{\mu})^2}{2\bar{\sigma}^2}}.
\end{align}
Accordingly, the joint CDF of $A_1, A_2, \dots, A_B$ can be formulated as
\begin{multline}\label{Bq5}
F_{A_1, A_2, \cdots, A_B}(r_1, \dots, r_B)=\\
\prod_{b=1}^B\left(\frac{1}{2}\mathrm{erf}\left(\frac{\gamma_b-\bar{\mu}}{\sqrt{2\bar{\sigma}^2}}\right)-\frac{1}{2}\mathrm{erf}\left(-\frac{\bar{\mu}}{\sqrt{2\bar{\sigma}^2}}\right)\right).
\end{multline}
Subsequently, the approximate outage probability can be found by substituting $\gamma_1=\cdots=\gamma_B=\sqrt{\gamma_{\rm th}}$ into the joint
CDF \eqref{Bq5}, which is given by
\begin{align}\label{Bq6}
\bar{\mathbb{P}}^{\mathrm{out}}=\left(\frac{1}{2}\mathrm{erf}\left(\frac{\sqrt{\gamma_{\rm th}}-\bar{\mu}}{\sqrt{2\bar{\sigma}^2}}\right)-\frac{1}{2}\mathrm{erf}\left(-\frac{\bar{\mu}}{\sqrt{2\bar{\sigma}^2}}\right)\right)^B.
\end{align}

\subsection{Throughput Optimization}
According to \eqref{eq9}, we know that the outage probability is an increasing function with respect to $R$. Thus, there exists a tradeoff between the target transmission rate $R$ and the coverage or non-outage probability $(1-\bar{\mathbb{P}}^{\mathrm{out}})$. Accordingly, given the expression of outage probability $\bar{\mathbb{P}}^{\mathrm{out}}$, the throughput can be maximized by optimizing $R$, which remains an open but interesting question. Subsequently, our aim is to optimize the  throughput of the considered RIS-FAS systems in the assumption of perfectly known CSI.

Based on the outage probability $\bar{\mathbb{P}}^{\mathrm{out}}$,  we first obtain the throughput of the MU, which is given by
\begin{align}\label{Eq1}
\bar{T}=R\left(1-\bar{\mathbb{P}}^{\mathrm{out}}\right).
\end{align}
Accordingly, the optimization problem can be formulated as
\begin{align}\label{Eq2}
\max_{R}\ \ \bar{T}\ \ \mbox{s.t.}\ R^{\min}\leq R \leq R^{\max},
\end{align}
where $R^{\min}$ and $R^{\max}$ denote the minimum rate and the maximum rate of the BS, respectively. It is difficult to directly solve Problem \eqref{Eq1}, since the convexity-concavity and monotonicity of the objective function $\bar{T}$ with respect to $R$ cannot be determined. To address this issue, we propose a gradient ascent algorithm to solve Problem \eqref{Eq1}, and obtain the locally optimal solution.

\subsubsection{Gradient Ascent  Algorithm}
In Problem \eqref{Eq1}, the objective function can be equivalently rewritten as
\begin{align}\label{Eq3}
\bar{T}=R\left(1-\left(1-Q\left(\frac{\sqrt{\gamma_{\rm th}}-\bar{\mu}}{\bar{\sigma}}\right)\right)^{B}\right),
\end{align}
where $Q(z)$ denotes the Gaussian $Q$-function, whose expression is
\begin{equation}\label{Eq4}
Q(z)=\frac{1}{\sqrt{2\pi}}\int_z^\infty e^{-\frac{t^2}{2}}dt.
\end{equation}

Taking the first-order derivative of $\bar{T}$, we have
\begin{align}
\Phi(R)&=\frac{\partial \bar{T}}{\partial R}\notag\\
&=1-\left(1-Q\left(\frac{\sqrt{\gamma_{\rm th}}-\bar{\mu}}{\bar{\sigma}}\right)\right)^{B}\nonumber\\
&~~~~-R\left(B\left(1-Q\left(\frac{\sqrt{\gamma_{\rm th}}-\bar{\mu}}{\bar{\sigma}}\right)\right)^{B-1}\right.\nonumber\\
&~~~~\left.\frac{1}{\sqrt{2\pi\bar{\sigma}^2}}e^{-\frac{(\sqrt{\gamma_{\rm th}}-\bar{\mu})^2}{2\bar{\sigma}^2}}\frac{\sigma^22^R\ln2}{2\sqrt{\sigma^2P(2^R-1)}}\right).
\end{align}

Denote that the optimization variable $R$ in the $t$-th iteration as $R^{t}$. Then, $R^{t+1}$ in the $(t + 1)$-th iteration is updated by
\begin{align}\label{Eq5}
R^{t+1} = R^{t}+\delta\Phi(R^{t}),
\end{align}
where $\delta$ is the step size of the gradient ascent algorithm. Thus, the solution to Problem \eqref{Eq7} is given by
\begin{align}
\min\{\max\{R^{\min}, R^o\}, R^{\max}\},
\end{align}
where $R^o$ is the solution by the gradient ascent algorithm.

\subsubsection{Bisection Search Method}
Due to the possible high computational complexity of the gradient ascent algorithm, we utilize the bisection search method as an alternative to reduce the complexity.  Specifically, when $0 \leq Q\left(z\right)\leq 1$, we have
\begin{align}\label{Eq5}
\left[1-Q\left(z\right)\right]^{B}\approx 1-BQ\left(z\right).
\end{align}
Therefore, $\bar{T}$ can be approximated by
\begin{align}\label{Eq6}
\bar{T}^{a}=&BRQ\left(\frac{\sqrt{\gamma_{\rm th}}-\bar{\mu}}{\bar{\sigma}}\right).
\end{align}
Despite these efforts, determining the convexity-concavity of $\bar{T}^{a}$ with respect to $R$ remains challenging. Thus, by adopting an approximation of the $Q$ function as $\frac{1}{2}e^{-\frac{z^2}{2}}$ for $z \geq 0$ \cite{BMakki14}, we have a further simplification of $\bar{T}^{a}$ as
\begin{align}\label{Eq6}
\bar{T}^{b} = \frac{1}{2}BR e^{-\frac{(\sqrt{\gamma_{\rm th}}-\bar{\mu})^2}{2\bar{\sigma}^2}}.
\end{align}
Then, Problem \eqref{Eq2} can be recast as
\begin{align}\label{Eq7}
\max_{R}\ \ \bar{T}^{b}\ \ \mbox{s.t.}\ R^{\min}\leq R \leq R^{\max}.
\end{align}

\emph{Lemma 1}: There exists a unique optimal $R$, denoted by $\bar{R}^o$, which maximizes $\bar{T}^{b}$, and is the solution to
\begin{equation}\label{Eq8}
\frac{\partial \bar{T}^{b}}{\partial R}=0.
\end{equation}

\begin{proof}
See Appendix A.
\end{proof}

From Lemma 1, $\bar{R}^o$ can be found by solving the equation \eqref{Eq8} by the bisection search method over the interval $[R^{\min}, R^{\max}]$. Considering the constraint on $R$, i.e.,
$R^{\min}\leq R \leq R^{\max}$, the optimal solution to Problem \eqref{Eq7} is
\begin{align}\label{bq4}
\min\{\max\{R^{\min}, \bar{R}^o\}, R^{\max}\}.
\end{align}

\section{CSI-Free Scheme}\label{sec:beyond}
In the previous section, we assumed that  BS  could perfectly acquire the CSI of the BS-RIS-MU link, and we proposed a CSI-based scheme with performance analysis and optimization design. However, the channel acquisition in FAS-RIS systems could be a challenging and heavy overhead task especially when the environment is dynamic. To address this, this section propose effective solutions for the harsh scenarios where the BS does not have the CSI of the BS-RIS-MU link. Consequently, we propose a CSI-free scheme with corresponding  performance analysis and optimizing design.
\subsection{Performance Analysis}
For the $b$-th submatrix of $\hat{\bm \Sigma}$, ${f}_k$ can be rewritten as
\begin{align}\label{Cq1}
& {f}_k\nonumber\\=&\sqrt{\frac{\alpha K}{K+1}}\bar{\mathbf{h}}_k\mathbf{\Phi}\mathbf{g}^H+ \mu_b\tilde{\mathbf{h}}_b\mathbf{\Phi}\mathbf{g}^H+ \sqrt{1-\mu_b^2}\mathbf{e}_k\mathbf{\Phi}\mathbf{g}^H\nonumber\\
=&\sum_{i=1}^M\bigg(\sqrt{\frac{\alpha K}{K+1}}\bar{h}_k^ie^{j\theta_i}g^i+\mu_b \tilde{h}_b^ie^{j\theta_i}g^i +\sqrt{1-\mu_b^2}e_k^ie^{j\theta_m}g^i\bigg).
\end{align}
Accordingly, we can observe that $\bar{h}^ie^{j\theta_m}g^i, i\in\mathcal{M}$ is the i.i.d RV. Besides,  when the number of the RIS reflection elements is sufficiently large and the phase $\phi_m$ follows the uniform distribution, we can apply the CLT to have
\begin{align}\label{Cq2}
\bar{\mathbf{h}}_k\mathbf{\Phi}\mathbf{g}^H \sim \mathcal{CN}(\rho,\eta^2),
\end{align}
where
\begin{align}
\rho&=M\cdot\mathbb{E}\left[\bar{h}^ie^{j\theta_i}g^i\right]=0,\label{Cq3}\\
\eta^2&=\sum_{i=1}^M\mathbb{E}\left[|h_0^ie^{j\theta_i}g^i-\rho|^2\right]=M\beta.\label{Cq4}
\end{align}

Similarly, $\tilde{\mathbf{h}}_b\mathbf{\Phi}\mathbf{g}^H$ is the sum of $\tilde{h}_b^ie^{j\theta_m}g^i, i\in\mathcal{M}$, and $\mathbf{e}_k\mathbf{\Phi}\mathbf{g}^H$ is the sum of $e_k^ie^{j\theta_m}g^i, i\in\mathcal{M}$. Thus, we have
\begin{align}\label{Cq5}
\tilde{\mathbf{h}}_b\mathbf{\Phi}\mathbf{g}^H &\sim \mathcal{CN}(\rho_b,\eta^2_b),\\
\mathbf{e}_k\mathbf{\Phi}\mathbf{g}^H &\sim \mathcal{CN}(\rho_e,\eta^2_e),
\end{align}
with
\begin{align}\label{Cq6}
\rho_b&=M\cdot\mathbb{E}\left[\tilde{h}_b^ie^{j\theta_i}g^i\right]=0,\\
\eta_b^2&=\sum_{i=1}^M\mathbb{E}\left[|\tilde{h}_b^ie^{j\theta_i}g^i-\rho_b|^2\right]=\frac{M\alpha\beta}{1+K},\\
\rho_e&=M\cdot\mathbb{E}\left[e_k^ie^{j\theta_i}g^i\right]=0,\\
\eta_e^2&=\sum_{i=1}^M\mathbb{E}\left[|e_k^ie^{j\theta_i}g^i-\rho_e|^2\right]=\frac{M\alpha\beta}{1+K}.
\end{align}

Moreover, the correlation coefficient between $\bar{\mathbf{h}}_k\mathbf{\Phi}\mathbf{g}^H$ and $\tilde{\mathbf{h}}_b\mathbf{\Phi}\mathbf{g}^H$ can be formulated as
\begin{align}\label{Cq7}
\rho_1 &= \frac{\mathbb{E}\left[\bar{\mathbf{h}}_k\mathbf{\Phi}\mathbf{g}^H\tilde{\mathbf{h}}_b\mathbf{\Phi}\mathbf{g}^H\right]-\mathbb{E}\left[\bar{\mathbf{h}}_k\mathbf{\Phi}\mathbf{g}^H\right]\mathbb{E}\left[\tilde{\mathbf{h}}_b\mathbf{\Phi}\mathbf{g}^H\right]}{\sqrt{\mathbb{V}\left[\bar{\mathbf{h}}_k\mathbf{\Phi}\mathbf{g}^H\right]}\sqrt{\mathbb{V}\left[\tilde{\mathbf{h}}_b\mathbf{\Phi}\mathbf{g}^H\right]}}\nonumber\\
&=\frac{\sum_{i=1}^M\sum_{k=1}^M\mathbb{E}\left[\bar{h}^ie^{j\theta_i}g^i\tilde{h}_b^ke^{j\theta_k}g^k\right]}{\eta\eta_b}=0.
\end{align}
Similarly,  the correlation coefficient between $\bar{\mathbf{h}}\mathbf{\Phi}\mathbf{g}^H$ and $\mathbf{e}_k\mathbf{\Phi}\mathbf{g}^H$ and that between $\mathbf{e}_k\mathbf{\Phi}\mathbf{g}^H$ and $\tilde{\mathbf{h}}_b\mathbf{\Phi}\mathbf{g}^H$ are both $0$. Since the sum of independent normal distributions still follows the normal distribution, we have
\begin{align}\label{Cq8}
\Upsilon = \sqrt{\frac{\alpha K}{K+1}}\bar{\mathbf{h}}_k\mathbf{\Phi}\mathbf{g}^H+\mu_b\tilde{\mathbf{h}}_b\mathbf{\Phi}\mathbf{g}^H\sim \mathcal{CN}\left(0,\hat{\sigma}^2\right),
\end{align}
where
\begin{align}\label{Cq9}
\hat{\sigma}^2=\frac{( K+\mu_b^2)M\alpha\beta}{K+1}.
\end{align}
Given $\Upsilon$, we know that
\begin{align}\label{Cq10}
\mathbf{h}_k\mathbf{\Phi}\mathbf{g}^H\sim \mathcal{CN}\left(\Upsilon,\check{\sigma}^2\right),
\end{align}
where
\begin{align}\label{Cq10-1}
\check{\sigma}^2=(1-\mu_b^2)\eta^2_e.
\end{align}

Letting $\Lambda_b \triangleq|\Upsilon|$, and given $\Lambda_b$, the PDF of $A_k$ can be expressed as \cite{KKWong21}
\begin{align}\label{Cq11}
f_{A_k | \Lambda_b}(r_k | r_b) = \frac{2r_k}{\check{\sigma}^2} e^{-\frac{r_k^2+ r_b^2}{\check{\sigma}^2}} I_{0}\left(\frac{2 r_k r_b}{\check{\sigma}^2}\right),
\end{align}
where $I_{0}(u)$ denotes the modified Bessel function of the first kind and order zero. Its series representation is given by \cite{ISGradshteyn07}
\begin{align}\label{Cq12}
I_0(z) = \sum_{k=0}^{\infty} \frac{z^{2k}}{2^{2k} k!\Gamma(k+1)},
\end{align}
where $\Gamma(k+1)=k!$.

Conditioned on $\Lambda_b$, it is evident that $A_1, \dots, A_N$ are mutually independent. This allows us to derive the joint PDF of $A_1, \dots, A_N$ as
\begin{align}\label{Cq13}
f_{A_1, \dots, A_N | \Lambda_b}(r_1, \dots, r_N| r_b)=&\prod_{k\in \mathcal{K}_b} \frac{2r_k}{\check{\sigma}^2} e^{-\frac{r_k^2+ r_b^2}{\check{\sigma}^2}}
I_{0}\left(\frac{2 r_k r_b}{\check{\sigma}^2}\right).
\end{align}
Building upon \eqref{Cq13}, the joint PDF of $\Lambda_b, A_1, \dots, A_N$ is
\begin{align}\label{Cq14}
f_{\Lambda_b, A_1, \dots, A_N}(r_b, \dots, r_N)=&\frac{2r_b}{\hat{\sigma}^2} e^{-\frac{r_b^2}{\hat{\sigma}^2}}\prod_{k\in \mathcal{K}_b} \frac{2r_k}{\check{\sigma}^2} e^{-\frac{r_k^2+ r_b^2}{\check{\sigma}^2}}\nonumber\\
&\times I_{0}\left(\frac{2 r_k r_b}{\check{\sigma}^2}\right).
\end{align}

According to \cite[(32)]{KKWong20}, the joint CDF of $\Lambda_b, A_1, \dots, A_N$ is given as
\begin{multline}\label{Cq15}
F_{\Lambda_b, A_1, A_2, \dots, A_N}(r_b, r_1, \dots, r_N)=\int_0^\infty \frac{2r_b}{\hat{\sigma}^2} e^{-\frac{r_b^2}{\hat{\sigma}^2}}\\
\times\prod_{k\in \mathcal{K}_b} \left[1-Q_1\left(\sqrt{\frac{2}{\check{\sigma}^2}}r_b,\sqrt{\frac{2}{\check{\sigma}^2}}r_k\right)\right]dr_b.
\end{multline}
Subsequently, the approximate outage probability can be found by substituting $r_1=\cdots=r_N=\sqrt{\gamma_{\rm th}}$ into the joint CDF in  \eqref{Cq15}, which is formulated as
\begin{align}\label{Cq16}
\tilde{\mathbb{P}}^{\mathrm{out}}=&\prod_{b=1}^B\int_0^\infty \frac{2r_b}{\hat{\sigma}^2} e^{-\frac{r_b^2}{\hat{\sigma}^2}}\nonumber\\
&\times\prod_{k\in \mathcal{K}_b} \left[1-Q_1\left(\sqrt{\frac{2}{\check{\sigma}^2}}r_b,\sqrt{\frac{2\gamma_{\rm th}}{\check{\sigma}^2}}\right)\right]dr_b.
\end{align}

The expression in \eqref{Cq16} involves integrals, requiring complex computations. Similar to the CSI-based scheme, we also employ the independent antennas equivalent model to eliminate the need for integrals in the approximations.

Specifically, when $\mu_b=1$, we have
\begin{align}\label{Dq1}
 {f}_k=\mathbf{h}_k\mathbf{\Phi}\mathbf{g}^H=&\sqrt{\frac{\alpha K}{K+1}}\bar{\mathbf{h}}_k\mathbf{\Phi}\mathbf{g}^H+\tilde{\mathbf{h}}_b\mathbf{\Phi}\mathbf{g}^H.
\end{align}
From \eqref{Cq2} to \eqref{Cq10}, we know that $\mathbf{h}_k\mathbf{\Phi}\mathbf{g}^H$ follows the complex Gaussian distribution by employing CLT, i.e.,
\begin{align}\label{Dq2}
\mathbf{h}_k\mathbf{\Phi}\mathbf{g}^H\sim \mathcal{CN}\left(0,1/\lambda_a\right),
\end{align}
where $\lambda_a = 1/(M\alpha\beta)$.

Thus, $|A_b|^2$ is an independent exponential distributed RV with parameter $\lambda_{a}$. The PDF of $|A_b|^2, b\in\mathcal{B}$ is given as
\begin{align}\label{Dq3}
f_{|A_b|^2}(z) = \left\{\begin{array}{ll}
\lambda_{a}e^{-\lambda_{a}z}, & z\geq 0, \\
0, & \mbox{otherwise}.
\end{array}\right.
\end{align}
Using \eqref{Dq3}, we can obtain the joint CDF of $|A_1|^2,\dots, |A_B|^2$ as
\begin{align}\label{Dq4}
F_{|A_1|^2, |A_2|^2, \dots, |A_B|^2}(r_1, \dots, r_B)=&\prod_{b=1}^B\left(1-e^{-\lambda_{a}r_b}\right).
\end{align}
As a result, the approximate outage probability can be found by substituting $r_1=\cdots=r_B=\gamma_{\rm th}$ into the joint CDF \eqref{Dq4}, which is given by
\begin{align}\label{Dq5}
\check{\mathbb{P}}^{\mathrm{out}}=\left(1-e^{-\lambda_{a}\gamma_{\rm th}}\right)^B.
\end{align}

\subsection{Throughput Optimization}
Similar to the CSI-based scheme, the optimization problem for the CSI-free scheme can be formulated as
\begin{align}\label{Fq1}
\max_{R}\ \ \check{T}\ \ \mbox{s.t.}\ 0\leq R \leq R^{\max},
\end{align}
where
\begin{align}\label{Fq2}
 \check{T}=R\left(1-\left(1-e^{-\lambda_{a}\gamma_{\rm th}}\right)^B\right).
\end{align}
It is also difficult to obtain the convexity-concavity and monotonicity of the objective function $\check{T}$ with respect to $R$. To address Problem \eqref{Fq1}, we respectively propose the partial gradient ascent algorithm and a closed-form solution method to solve Problem \eqref{Eq1}, where the use of the partial gradient ascent algorithm means that the gradient ascent algorithm is only used in the partial interval of the optimization variable $R$.

\subsubsection{Partial Gradient Descent Algorithm} To provide the accuracy and reduce the computational complexity simultaneously, we propose the partial gradient ascent algorithm. Let $x=2^R-1(x\geq 0)$, i.e., we have $R=\log_2(1+x)$. The objective function can be rewritten as
\begin{align}\label{Fq3}
 \check{T}=&\log_2(1+x)\left(1-\left(1-e^{-\Gamma x}\right)^B\right),
\end{align}
where $\Gamma = \frac{\lambda_a\sigma^2}{P}$.

Taking the first-order derivative and the second-order derivative of $ \check{T}$ with respect to $x$, we have
\begin{align}\label{Fq4}
\frac{\partial \check{T}}{\partial x}=&\frac{1}{\ln2(1+x)}\left(1-\left(1-e^{-\Gamma x}\right)^B\right)\nonumber\\
&-\log_2(1+x)B\Gamma\left(1-e^{-\Gamma x}\right)^{B-1}e^{-\Gamma x},\\
\label{Fq5}\frac{\partial^2 \check{T}}{\partial x^2}=&-\frac{1}{\ln2(1+x)^2}\left(1-\left(1-e^{-\Gamma x}\right)^B\right)\nonumber\\
&-\frac{2}{\ln2(1+x)}B\Gamma\left(1-e^{-\Gamma x}\right)^{B-1}e^{-\Gamma x}-\log_2(1+x)\nonumber\\
&\times B\Gamma^2\left(1-e^{-\Gamma x}\right)^{B-2}e^{-\Gamma x}\left(Be^{-\Gamma x}-1\right).
\end{align}
From \eqref{Fq5}, we know that $\frac{\partial^2 \check{T}}{\partial R^2}\leq0$ when $Be^{-\Gamma x}-1\geq0$, which can be transformed as $x\leq\frac{\ln B}{\Gamma}$. Therefore, we can obtain that $\check{T}$ is a concave function with respect to $x$, when $x\leq\frac{\ln B}{\Gamma}$. We define $\check{x}=\frac{\ln B}{\Gamma}$. To obtain $x$ that maximizes $\check{T}$ for $x\in\left[0,\check{x}\right]$, we have the following lemma.

\emph{Lemma 3}: For $x\in\left[0,\check{x}\right]$, if $\left.\frac{\partial \check{T}}{\partial x}\right|_{x=\check{x}}\geq0$, $x$ maximizing $\check{T}$ is $\frac{\ln B}{\Gamma}$. If $\left.\frac{\partial \check{T}}{\partial x}\right|_{x=\check{x}}<0$, $x$ maximizing $\check{T}$ is the solution to
\begin{equation}\label{Fq6}
\frac{\partial \check{T}}{\partial x}=0,
\end{equation}
which can be found by employing the bisection search method.

\begin{proof}
See Appendix B.
\end{proof}

According to Lemma 3, we define $x$ that maximizes $\check{T}$ as $x^\dagger$, and $R$ that maximizes $\check{T}$ can be expressed as
\begin{equation}\label{Fq7}
R^\dagger=\log_2(1+x^\dagger).
\end{equation}
For $x>\frac{\ln B}{\Gamma}$, because the convexity-concavity and monotonicity of the objective function  $\check{T}$ with respect to $R$ cannot be determined, we use the gradient ascent algorithm to solve \eqref{Fq1}. Assume the optimization variable $x$ in the $t$-th iteration is $x^{t}$. Then, $x^{t+1}$ in the $(t + 1)$-th iteration is given by
\begin{align}\label{Fq8}
x^{t+1} = x^{t}+\delta\left.\frac{\partial \check{T}}{\partial x}\right|_{x=x^{t}},
\end{align}
where $\delta$ is the step size of the gradient ascent algorithm. We define the solution of $x$ obtained by the gradient ascent algorithm as $x^\ddagger$, and then the corresponding $R$ is
\begin{equation}\label{Fq8}
R^\ddagger=\log_2(1+x^\ddagger).
\end{equation}
Thus, the optimal $R$ can be obtained by
\begin{align}\label{Fq9}
R^\ast = \left\{\begin{array}{ll}
R^\dagger, & \check{T}(R^\dagger)\geq \check{T}(R^\ddagger), \\
R^\ddagger, & \mbox{otherwise}.
\end{array}\right.
\end{align}
The solution to Problem \eqref{Fq1} is
\begin{align}
\min\{\max\{R^{\min}, R^\ast\}, R^{\max}\}.
\end{align}

\subsubsection{Closed-Form Solution}
Though the partial gradient ascent algorithm can reduce the computational complexity, the complexity is still high when the optimal $R$ is much larger than $\log_2\left(1+\check{x}\right)$. To further reduce the complexity,  we propose a closed-form solution method. We use the approximation in \eqref{Eq5} to approximate $ \check{T}$, which can be expressed as
\begin{align}\label{Fq10}
 \check{T}^{a}=&BRe^{-\lambda_{a}\gamma_{\rm th}}.
\end{align}
Therefore, Problem \eqref{Fq1} is recast as
\begin{align}\label{Fq11}
\max_{R}\ \ \check{T}^{a}\ \ \mbox{s.t.}\ R^{\min}\leq R \leq R^{\max}.
\end{align}

\emph{Lemma 4}: The objective function of the Problem \eqref{Fq11} $ \check{T}^{a}$  first increases and then decreases as $R$ increases. The closed-form solution for $\frac{\partial \check{T}^{a}}{\partial R}=0$ is provided by
\begin{align}\label{Fq12}
\check{R}^\star=\frac{1}{\ln2}\mathcal{W}\left(\frac{P}{\lambda_a\sigma^2} \right),
\end{align}
where $\mathcal{W}(z)$ represents the Lambert $\mathcal{W}$ function.

\begin{proof}
See Appendix C.
\end{proof}

Moreover, we also obtain another closed-form solution as
\begin{align}\label{Fq13}
\bar{R}^\star= \log_2\left(1+\check{x}\right).
\end{align}
Thus, the optimal $R$ can be obtained by
\begin{align}\label{Fq14}
R^\star= \left\{\begin{array}{ll}
\check{R}^\star, & \check{T}(\check{R}^\star)\geq \check{T}(\bar{R}^\star), \\
\bar{R}^\star, & \mbox{otherwise}.
\end{array}\right.
\end{align}
Finally, the optimal solution to Problem \eqref{Fq11} becomes
\begin{align}\label{Fq15}
\min\{\max\{R^{\min}, R^\star\}, R^{\max}\}.
\end{align}

\section{Numerical Results}\label{sec:results}
In our simulations, we consider a three-dimensional (3D) coordinate system. The BS, RIS, and MU are located at $(0, 0, 0)~{\rm m}$, $(10, 10, 5)~{\rm m}$, and $(50, 0, 0)~{\rm m}$, respectively. The channels from the BS to the MU are modeled as distance-dependent flat Rician fading channels, where the large-scale path loss exponent is $2.2$ and the Rician factor is one \cite{BWei24}. The angles in the LoS channels are generated randomly from $[0, 2\pi]$. The noise power is $\sigma^2=-104~{\rm dBm}$. We assume that a parameter value of $W=5$. We also assume that the correlation coefficient of each block in \eqref{eq11} is identical, i.e., $\mu_{1}^2=\cdot\cdot\cdot =\mu_{b}^2=\cdot\cdot\cdot=\mu_{B}^2=0.97$ \cite{PR24}. The Monte Carlo simulations are obtained based on the correlation coefficient matrix in \eqref{eq5} with $100000$ time average.

\begin{figure}[t]
    \centering
    \begin{minipage}[b]{0.48\linewidth}
        \centering
        \includegraphics[width=\textwidth]{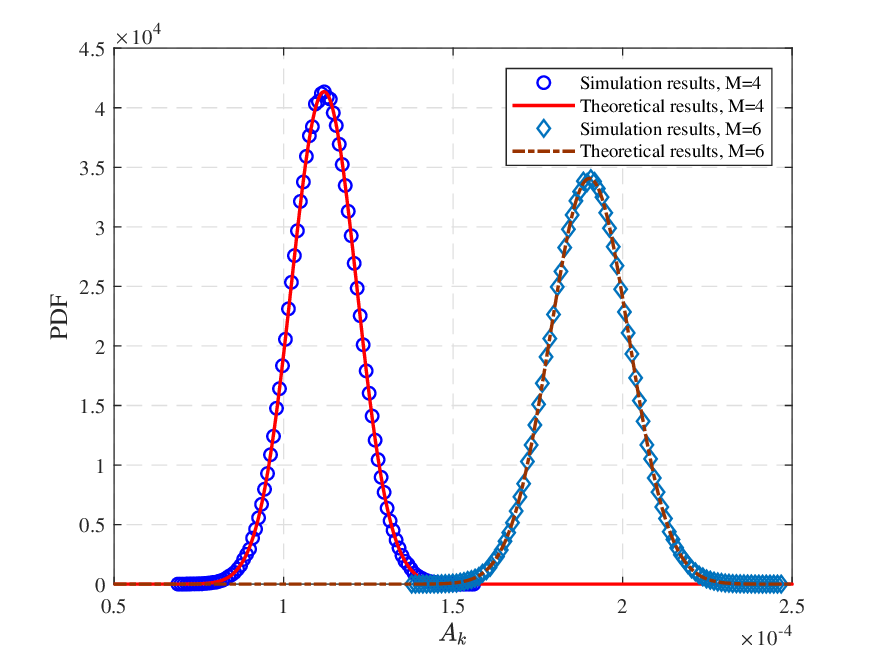}
        \caption{PDFs of the $A_k$ in the CSI-based case.  }
        \label{fig:pdf1}
    \end{minipage}
    \hfill
    \begin{minipage}[b]{0.48\linewidth}
        \centering
        \includegraphics[width=\textwidth]{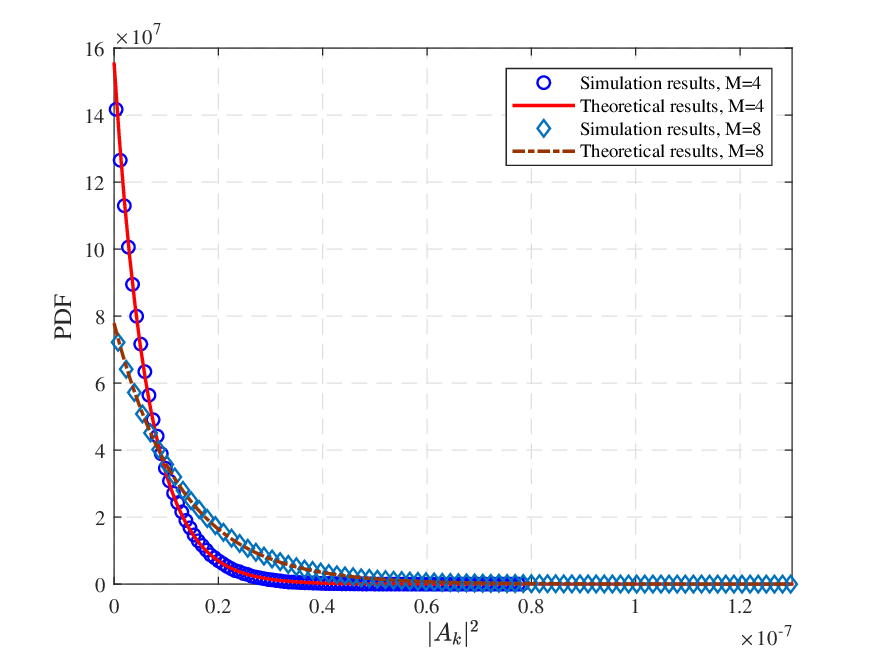}
        \caption{PDFs of  $|A_k|^2$ in the CSI-free scheme. }
        \label{fig:pdf2}
    \end{minipage} 
\end{figure} 

In Fig.~\ref{fig:pdf1}, we illustrate the PDFs of $A_k$ in the CSI-based case. According to \eqref{Aq2}, it is known that $A_k$ follows the normal distribution. In the legend, ``Simulation results" is the result computed based on Monte Carlo, and ``Theoretical results'' denotes the results based on theoretical distribution $\frac{1}{\sqrt{2\pi \sigma^2_k}} e^{-\frac{(r_k-\mu_k)^2}{2\sigma^2_k}}$. It can be observed from Fig.~\ref{fig:pdf1} that the theoretical results match the Monte Carlo results well.

In Fig.~\ref{fig:pdf2}, we show the PDFs of $|A_k|^2$ in the CSI-free scheme. From \eqref{Cq2} to \eqref{Cq7}, it is understood that $A_k$ follows the complex Gaussian distribution. Thus, $|A_k|^2$  follows the exponential distribution. It is shown again that the theoretical results match the Monte Carlo results well.
\begin{figure}[t]
    \centering
    \begin{minipage}[b]{0.48\linewidth}
        \centering
        \includegraphics[width=\textwidth]{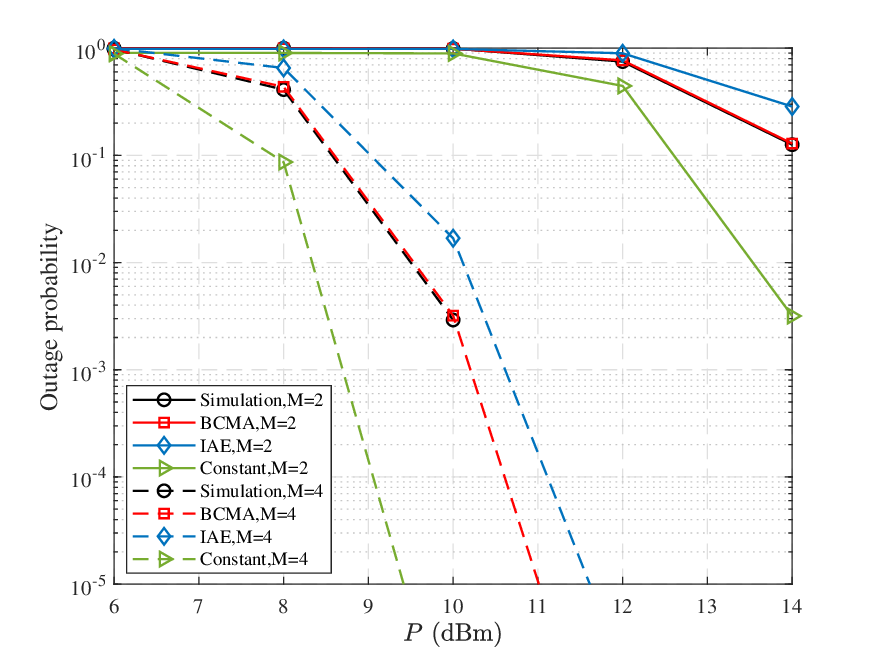}
        \caption{Outage probability versus $P$ in the CSI-based scheme, where $N=100$ and $R=2~{\rm bps/Hz}$.}
        \label{fig:oVSp1}
    \end{minipage}
    \hfill
    \begin{minipage}[b]{0.48\linewidth}
        \centering
        \includegraphics[width=\textwidth]{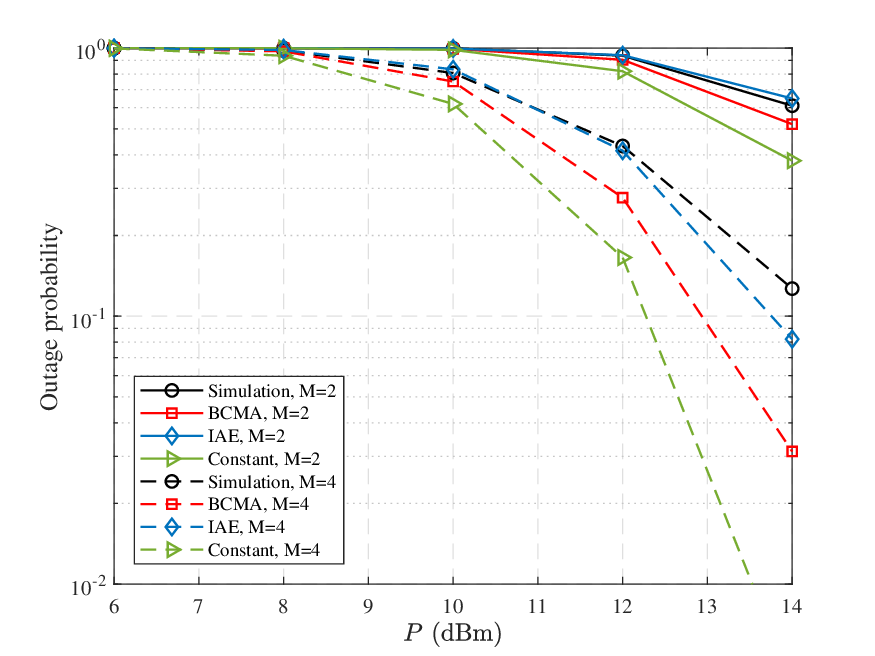}
        \caption{Outage probability versus $P$ in the CSI-free scheme, where $N=100$ and $R=2~{\rm bps/Hz}$.}
        \label{fig:oVSp2}
    \end{minipage}
\end{figure}

In Figs.~\ref{fig:oVSp1} and \ref{fig:oVSp2}, we  examine the impact of the transmit power of the BS, i.e., $P$, on the outage probability in the CSI-based scheme and the CSI-free scheme, respectively. In the legend, ``Simulation" refers to the simulated outage probability, while ``BCMA" denotes the theoretical analysis based on the block-correlation matrix approximation model. ``IAE" denotes the theoretical analysis based on the independent antenna equivalent model, and ``Constant" denotes the theoretical analysis based on the constant-correlation matrix model, where the correlation coefficient matrix is given by \cite{KKWong21}
\begin{equation}
\mathbf{\Sigma}=
   \left[ \begin{array}{cccc}
1& \mu^2 &\cdots  &  \mu^2  \\
\mu^2 & 1 &\cdots  &  \mu^2 \\
 \vdots &  \vdots &\ddots & \vdots \\
\mu^2 &\mu^2 & \cdots &1 \\
\end{array}\right].
\end{equation}
Observing from the results in Figs.~\ref{fig:oVSp1} and \ref{fig:oVSp2}, it can be observed that the outage probabilities of all schemes decrease as $P$ increases. Besides, we also find that the larger $M$, the smaller outage probabilities of all schemes for a given $P$. This is because the larger $M$, the larger DoFs provided by the RIS. In Fig.~\ref{fig:oVSp1}, it is evident that the lines of ``BCMA" closely match the simulated results, and the approximate accuracy of ``BCMA" is better than that of ``IAE". This is because the block-correlation matrix approximation model can effectively capture the spectrum of the true correlation matrix in \eqref{eq5}, but the independent antenna equivalent model cannot in the CSI-based case. On the contrary, in Fig.~\ref{fig:oVSp2}, the approximate accuracy of ``BCMA" is worse than that of ``IAE". Thus, ``BCMA" and ``IAE" are both effective approximation models. Moreover, in Fig.~\ref{fig:oVSp1} and Fig.~\ref{fig:oVSp2}, the approximate accuracy of ``Constant" is always worst and the ``IAE" is always the upper bound of the ``BCMA", which are consistent with Remark 1.

\begin{figure}[t]
    \centering
    \begin{minipage}[b]{0.48\linewidth}
        \centering
        \includegraphics[width=\textwidth]{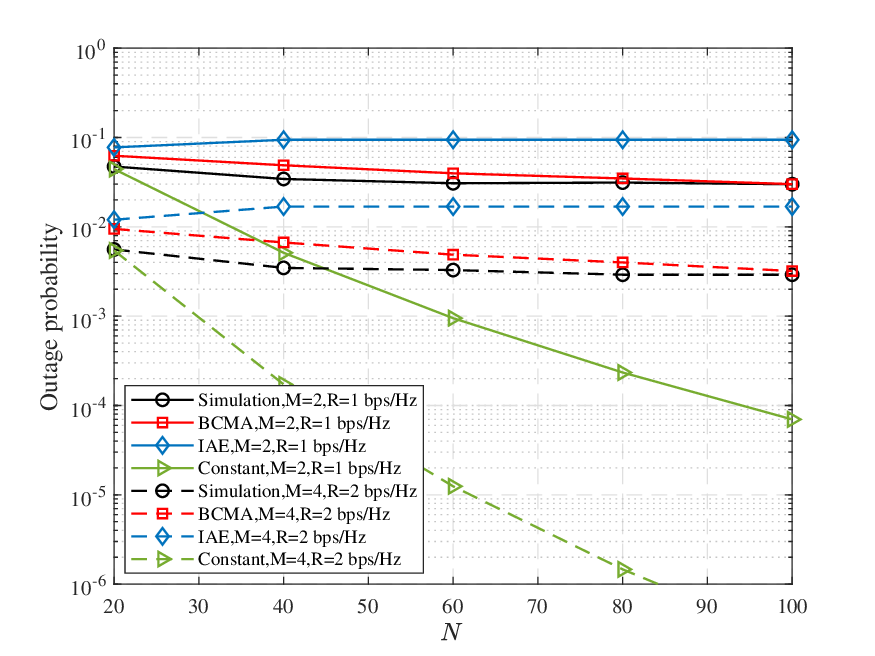}
       \caption{Outage probability versus $N$ in CSI-based case, where $P=10~{\rm dBm}$.}\label{fig:oVSn1}
    \end{minipage}
    \hfill
    \begin{minipage}[b]{0.48\linewidth}
        \centering
        \includegraphics[width=\textwidth]{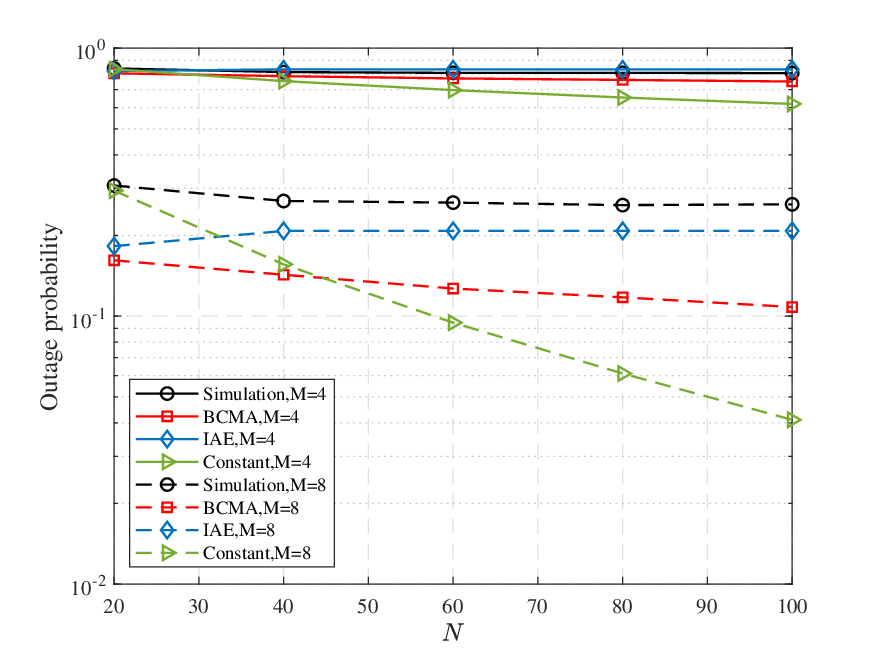}
       \caption{Outage probability versus $N$ in CSI-free case, where $P=10~{\rm dBm}$ and $R=2~{\rm bps/Hz}$.}\label{fig:oVSn2}
    \end{minipage}
\end{figure}

In Fig.~\ref{fig:oVSn1}, we demonstrate the impact of $N$ on the outage probability in the CSI-based case, where $P=10~{\rm dBm}$. As is seen from Fig.~\ref{fig:oVSn1}, the outage probabilities of all schemes decrease as $N$ increases except for the ``IAE". This is because the larger $N$, the larger the SNR of the user.  The outage probabilities of ``IAE" increase as $N$ increases when $N$ changes from $20$ to $40$. This is because the outage probabilities of ``IAE" mainly depend on the number of the blocks in $\hat{\Sigma}$, and the number of the blocks for $N=20$ is larger than that for $N=40$. From Fig.~\ref{fig:oVSn1}, we can also find that the larger $N$, the more accurate  the outage performance of ``BCMA". This is because the block-correlation matrix approximation is based on the asymptotic statistical results for larger $N$. Also, it can be observed that the impact of $M$ is significant on the outage probabilities of all schemes, and the outage probabilities for $M=4$ are smaller than those for $M=2$, even if $R=2~{\rm bps/Hz}$ for $M=4$ is twice as much as that for $M=2$. Similar with the results in Figs.~\ref{fig:oVSp1} and \ref{fig:oVSp2}, the approximate accuracy of ``Constant" is worst. In Fig.~\ref{fig:oVSn2}, we show the impact of $N$ on the outage probability without CSI, where $P=10~{\rm dBm}$ and $R=2~{\rm bps/Hz}$. From Figs.~\ref{fig:oVSn1} and \ref{fig:oVSn2}, we can find that the outage performance with CSI is better than that without CSI. Specifically, when $N=100$, $P=10~{\rm dBm}$ and $R=2~{\rm bps/Hz}$, the outage probability of ``Simulation" with CSI and $M=4$ is about $0.0031$, while the outage probability of ``Simulation" without CSI and with $M=4$ is about $0.8337$ and the outage probability of ``Simulation" without CSI and with $M=8$ is about $0.2615$.
\begin{figure}[t]
    \centering
    \begin{minipage}[b]{0.48\linewidth}
        \centering
        \includegraphics[width=\textwidth]{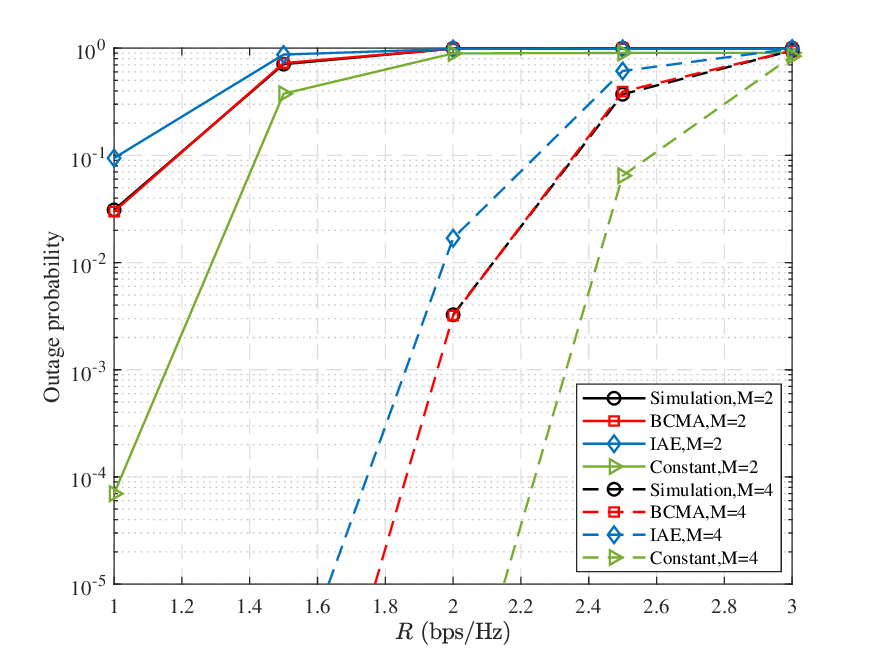}
\caption{Outage probability versus $R$ in CSI-based case, where $N=100$ and $P=10~{\rm dBm}$.}\label{fig:oVSr1}
    \end{minipage}
    \hfill
    \begin{minipage}[b]{0.48\linewidth}
        \centering
        \includegraphics[width=\textwidth]{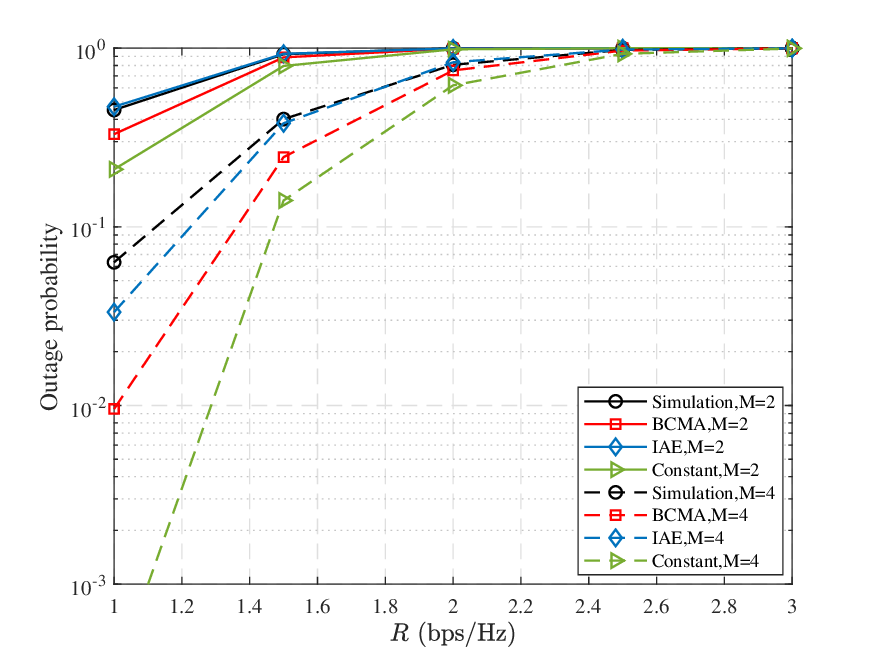}
\caption{Outage probability versus $R$ in CSI-free case, where $N=100$ and $P=10~{\rm dBm}$.}\label{fig:oVSr2}
    \end{minipage}
\end{figure}

In Figs.~\ref{fig:oVSr1} and \ref{fig:oVSr2}, the results are provided for the outage probability across various values of $R$ with $N=100$ and $P=10~{\rm dBm}$.  Observing from Fig.~\ref{fig:oVSr1} and Fig.~\ref{fig:oVSr2}, we can see that the outage probabilities of all schemes increase as $R$ increases. In Fig.~\ref{fig:oVSr1}, it can be seen that the lines of ``BCMA" closely match the simulated results, which demonstrates that the block-correlation matrix approximation model is very effective for the CSI-based case. In Fig.~\ref{fig:oVSr2}, we can also see that the outage performance gap between  ``Simulation" and ``IAE" is smaller than that between ``Simulation" and ``BCMA".

\begin{figure}[h]
\centering
\includegraphics[width=3.4in]{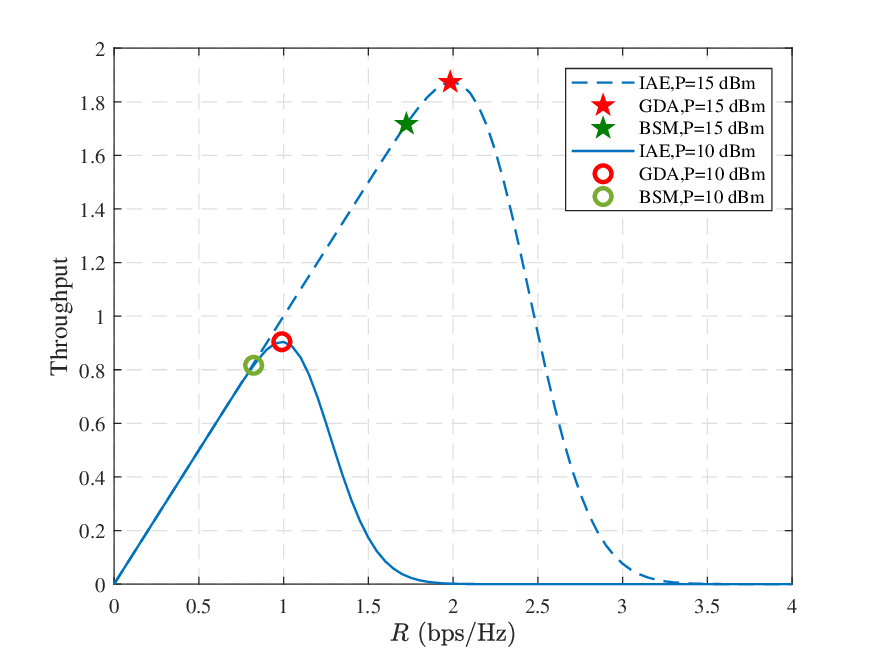}
\caption{Throughput versus $R$ in CSI-based scheme, where $M=2$ and $N=100$.}\label{fig:tVSr1}
\end{figure}

\begin{figure}[h]
\centering
\includegraphics[width=3.4in]{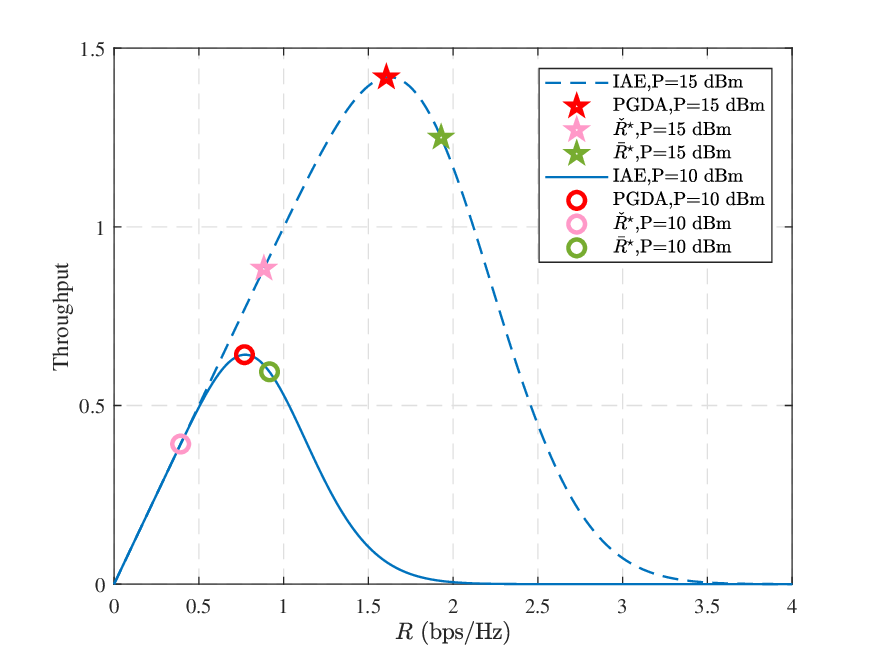}
\caption{Throughput versus $R$ in CSI-free case, where $M=2$ and $N=100$.}\label{fig:tVSr2}
\end{figure}

In Fig.~\ref{fig:tVSr1}, the results are provided for the throughput based on the independent antennas equivalent mode across various values of $R$ with $M=2$ and $N=100$ in the CSI-based case. In the legend, ``IEA" means the theoretical analysis based on the independent antenna equivalent model;  ``GDA" denotes the solution provided by the gradient ascent algorithm, while ``BSM" denotes the solution provided by the bisection search method. Observing from Fig.~\ref{fig:tVSr1}, we can see that ``GDA" can obtain near optimal solution, whereas ``BSM" cannot. This is because the ``BSM" uses the approximate objective function. The advantage of ``BSM" is of low computational complexity, which makes at the expense of accuracy. It can also be seen from the results in Fig.~\ref{fig:tVSr1} that the throughput increases as $P$ increases. Fig.~\ref{fig:tVSr2} shows the throughput based on the independent antenna equivalent model across various values of $R$ given $M=2$ and $N=100$ and without CSI. In the legend, ``PGDA" denotes the solution provided by the partial gradient ascent algorithm; ``$\check{R}^\star$" denotes the solution in \eqref{Fq12}, while ``$\bar{R}^\star$" denotes the solution in \eqref{Fq13}. Similarly, ``GDA" can obtain a near-optimal solution. Besides, the solution of ``$\check{R}^\star$" is worse than ``$\bar{R}^\star$".  Moreover, we can find that ``$\bar{R}^\star$" is larger than the optimal $R$. In this situation, the optimal $R$ can be found by using bisection search alone, without the need for the gradient ascent algorithm, which can reduce the computational complexity significantly.

\begin{figure}[h]
\centering
\includegraphics[width=3.4in]{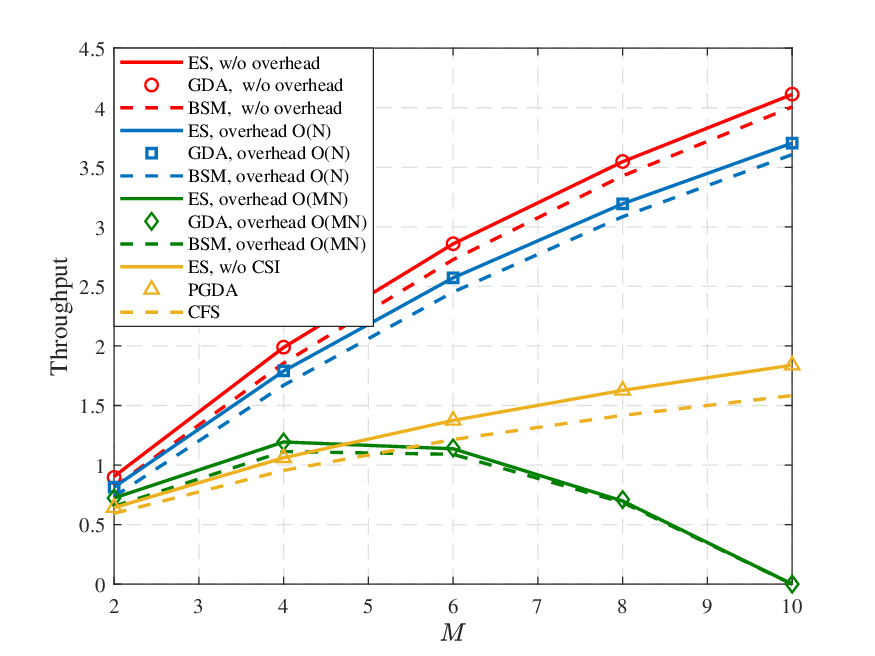}
\caption{Throughput versus $M$,  where $P=10$ dBm and $N=100$.}\label{fig:tVSm2}
\end{figure}

\begin{figure}[h]
\centering
\includegraphics[width=3.4in]{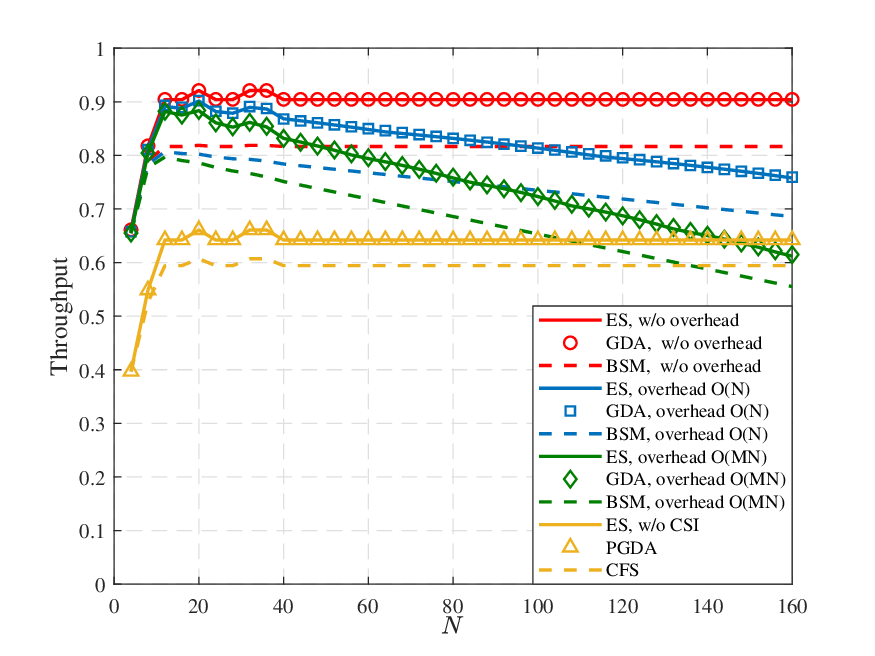}
\caption{Throughput versus $N$,  where $P=10$ dBm and $M=2$.}\label{fig:tVSn2}
\end{figure}

\begin{figure}[h]
\centering
\includegraphics[width=3.4in]{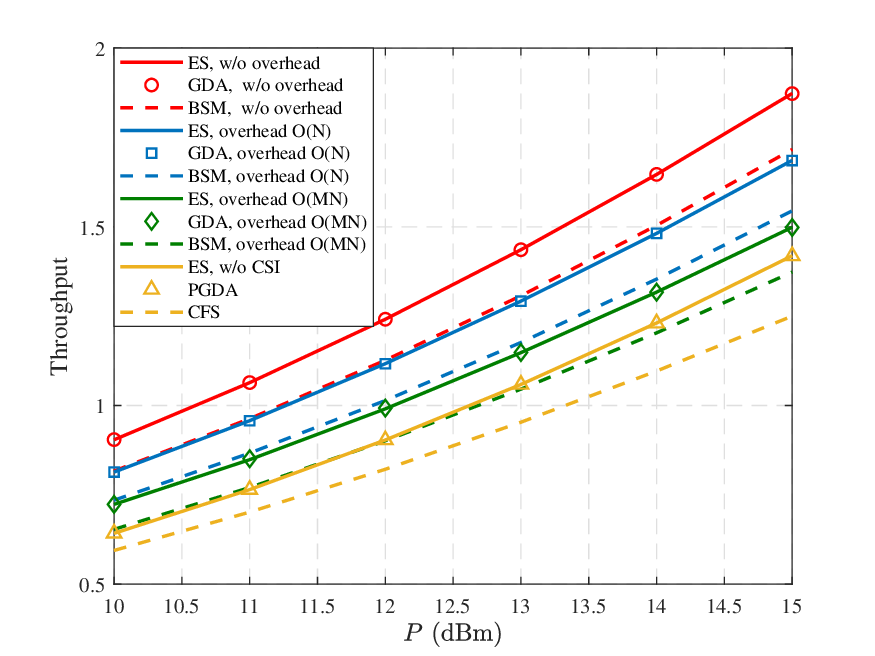}
\caption{Throughput versus $P$,  where $M=2$ and $N=100$.}\label{fig:tVSp1}
\end{figure}

\begin{figure}[h]
\centering
\includegraphics[width=3.4in]{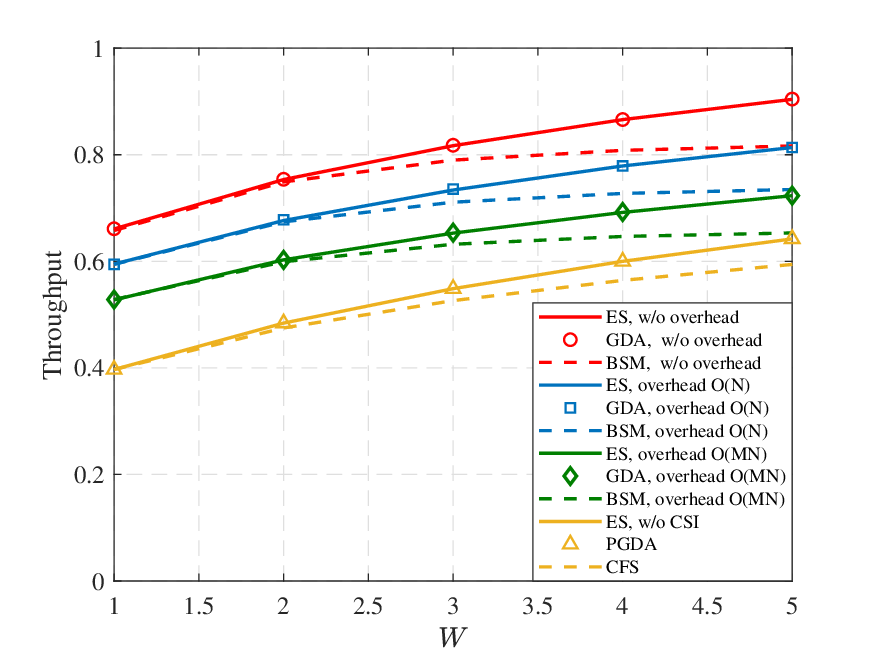}
\caption{Throughput versus $W$,  where $P=10$ dBm, $M=2$ and $N=100$.}\label{fig:tVSw2}
\end{figure}

In Fig.~\ref{fig:tVSm2}, Fig.~\ref{fig:tVSn2}, Fig.~\ref{fig:tVSp1}, and Fig.~\ref{fig:tVSw2}, we compare the throughput of the CSI-based scheme and CSI-free scheme. To provide a fair comparison in the practical applications, here we further consider the rate loss caused by channel estimation overhead. In each channel coherence time, assuming that there are $S$ time slots, if channel estimation consumes $\tau$ time slots, only $S-\tau$ time slots can be left for data transmission. Then, the overhead-aware outage probability can be expressed as
\begin{align}\label{eq87}
\mathbb{P}^{\mathrm{out}}&=\mathbb{P}\left(\left(1-\frac{\tau}{\Omega}\right)\log_2(1+\gamma)<R\right).
\end{align}
We assume that $\Omega=1000$ \cite{QTao21}. In CSI-based cases, if we neglect the channel estimation overhead and let $\tau=0$, it is actually an ideal the upper bound which will be denoted as ``w/o overhead'' in the legend. To quantify the pilot overhead of CSI-based cases in FAS-RIS systems, we consider two simple cases. First, we consider a relatively ideal case by neglecting the impact of RIS on channel estimation. In this case, we only consider the channel estimation overhead caused by FAS, so that we have $\tau=N$. This case is referred to as ``overhead O(N)'' in the legend. Secondly, further taking into consideration the channel estimation difficulty from RIS, we consider using the on-off scheme so that the overall overhead could be $\tau=NM$, which is referred to as ``overhead O(NM)'' in the legend. For CSI-free cases, we do not need to estimate CSI and therefore we have $\tau=0$.

In Fig.~\ref{fig:tVSm2}, we examine   the throughput based on the independent antenna equivalent mode across various values of $M$ with $P=10$ dBm and $N=100$. In the legend, ``ES" and ``ES, w/o CSI" mean that the optimal solutions obtained by the exhaustive search method in with CSI case and CSI-free case, respectively. Observing from Fig.~\ref{fig:tVSm2}, we can see the ``w/o overhead'' schemes achieves the larger throughput than other schemes, which is the upper bound of throughput. We can also observe that the curves of the ``overhead O(NM)'' schemes   first increase but then decrease. With considering channel estimation overhead, the CSI-free scheme could even outperform the CSI-based scheme. This is because even though CSI-based design can achieve a better performance, it is at the cost of overhead. When $M, N$ are large, most of the time slots are used for transmitting pilots, and only limited time is left for data transmission. In that case, the performance is dominated by pilot overhead, and the gain from using CSI is overwhelmed. This demonstrates the effectiveness of the pilot-free scheme, i.e., CSI-free scheme, in the cases of short-channel coherence time scenarios and heavy-overhead scenarios.

In Fig.~\ref{fig:tVSn2}, we study that the throughput based on the independent antenna equivalent mode across various values of $N$ with $P=10$ dBm and $M=2$. Observing from Fig.~\ref{fig:tVSn2}, we can see the curves of all scheme are fluctuated when $N$ varies from 20 to 40, since the number of diagonal matrix $B$ fluctuated. The curves of ``w/o overhead'' schemes, the ``ES, w/o CSI" scheme, the ``PGDA'' scheme, and the ``CFS'' scheme are flat when $N$ is larger than 40 due to the constant $B$. The curves of the ``overhead O(N)''  schemes and the ``overhead O(NM)''  schemes  are first increase and then decrease. This is because the benefits brought by the number of the ports is significant when $N$ is small, whereas the overhead becomes dominant when $N$ is large.

In Fig.~\ref{fig:tVSp1}, we show that the throughput based on the independent antenna equivalent mode across various values of $P$ with $M=2$ and $N=100$. Observing from Fig.~\ref{fig:tVSp1}, we can see that ``GDA" and ``PGDA" can obtain near optimal solution in CSI-based case and  CSI-free case, respectively. From Fig.~\ref{fig:tVSp1}, we can find that the throughput of two cases increasing with the increasing of $P$. Besides, we also can observe from Fig.~\ref{fig:tVSp1} that the throughput gap between ``GDA" and ``BSM", and that between ``PGDA" and ``CFS" increase as $P$ increases. This is because the smaller $Q\left(z\right)$ in \eqref{Eq5}, more accurate the approximation in \eqref{Eq5}, and $Q\left(\frac{\sqrt{\gamma_{th}}-\bar{\mu}}{\bar{\sigma}}\right)$ increases as $P$ increases. The same reason for CSI-free case.

In Fig.~\ref{fig:tVSw2}, we investigate that the throughput based on the independent antenna equivalent mode across various values of $W$ with $P=10$ dBm, $M=2$, and $N=100$. Observing from Fig.~\ref{fig:tVSw2}, we can see that the ``w/o overhead'' schemes achieve  the larger throughput than other schemes. Furthermore, we also can observe from Fig.~\ref{fig:tVSw2} that the throughput gap between ``GDA" and ``BSM", and that between ``PGDA" and ``CFS" increase  with the increasing of $W$. This is because when the value of $W$ increases, the statistical dependency among the channel of each port decreases, and $B$ becomes larger.  Besides, the larger $B$ in \eqref{Eq5},  the worse accurate the approximation in \eqref{Eq5}.

\section{Conclusion}\label{sec:conclude}
In this paper, we investigated the RIS-FAS systems in which the BS transmits the information signals to the FAS-enabled MU with the aid of RIS. We have proposed a comprehensive framework consisting of a CSI-based scheme and a CSI-free scheme. For the two schemes, we introduced two approximation models to obtain the corresponding outage probabilities of the systems. Using the approximate outage probabilities, we maximized the system throughput using the two schemes by designing the transmission rate of the BS. For the CSI-based scheme, we proposed the gradient ascent algorithm and the bisection search method to obtain a near-optimal solution and a locally optimal solution. For the CSI-free scheme, we proposed the partial gradient ascent algorithm and a closed-form solution. Our results demonstrated that the effectiveness of our proposed framework for the RIS-FAS system.

\appendices
\section{Proof of Lemma 1}
To prove Lemma 1, we need the following lemma, whose proof can be found in \cite{Boyd}.

\emph{Lemma 2}: Define $f(x)$ as a quadratic differentiable function on the convex set $\mathbf{dom} f \subset \mathbb{R}$. If $f(x)$ satisfies $f'(x)=0\Rightarrow f''(x)<0$, i.e., $x\in \mathbf{dom} f$ that satisfies the first-order derivative being zero makes the second-order derivative be smaller than $0$, $f(x)$ is a quasiconcave function. $\hfill\blacksquare$

Letting $x=2^R-1(x\geq 0)$, we have $R=\log_2(1+x)$. The objective function $\bar{T}^{b}$ can be rewritten as
\begin{align}
F(x)=\frac{B}{2}\log_2(1+x)e^{-\frac{(\kappa\sqrt{x}-\bar{\mu})^2}{2\bar{\sigma}^2}},
\end{align}
where $\kappa =\sqrt{ \frac{\sigma^2}{P}}$.

Taking the first-order derivative of $F(x)$ with respect to $x$, after mathematical manipulations, we have
\begin{align}\label{Gq1}
\frac{\partial F(x)}{\partial x}=\frac{B}{2}(h(x)-g(x)),
\end{align}
where
\begin{align}
h(x)=&\frac{1}{\ln2(1+x)}e^{-\frac{(\kappa\sqrt{x}-\bar{\mu})^2}{2\bar{\sigma}^2}},\\
g(x)=&\log_2(1+x)e^{-\frac{(\kappa\sqrt{x}-\bar{\mu})^2}{2\bar{\sigma}^2}}\frac{\kappa(\kappa\sqrt{x}-\bar{\mu})}{2\bar{\sigma}^2\sqrt{x}}.
\end{align}

Define $\hat{x}$ that satisfies $\left.\frac{\partial F(x)}{\partial x}\right|_{x=\hat{x}}=0$. Because $h(\hat{x})\geq0$, we know that $g(\hat{x})\geq0$, i.e., $\kappa\sqrt{\hat{x}}-\bar{\mu}\geq0$.
Then, taking the second-order derivative of $F(x)$ with respect to $x$, we have

\begin{align}\label{Gq7}
\frac{\partial^2 F(x)}{\partial x^2} = &\frac{B}{2}\left(-\frac{1}{\ln2(1+x)^2}e^{-\frac{(\kappa\sqrt{x}-\bar{\mu})^2}{2\bar{\sigma}^2}}-f(x)\right.\nonumber\\
&\left.+l(x)-\log_2(1+x)e^{-\frac{(\kappa\sqrt{x}-\bar{\mu})^2}{2\bar{\sigma}^2}}\frac{\kappa\bar{\mu}}{4\bar{\sigma}^2x^{\frac{3}{2}}}\right),
\end{align}
where
\begin{align}
f(x)=&2h(x)\frac{\kappa(\kappa\sqrt{x}-\bar{\mu})}{2\bar{\sigma}^2\sqrt{x}},\\
l(x)=&\log_2(1+x)e^{-\frac{(\kappa\sqrt{x}-\bar{\mu})^2}{2\bar{\sigma}^2}}\left(\frac{\kappa(\kappa\sqrt{x}-\bar{\mu})}{2\bar{\sigma}^2\sqrt{x}}\right)^2.
\end{align}
Since $h(\hat{x})=g(\hat{x})$, we can obtain that
\begin{align}
f(\hat{x})=&2\log_2(1+\hat{x})e^{-\frac{(\kappa\sqrt{\hat{x}}-\bar{\mu})^2}{2\bar{\sigma}^2}}\left(\frac{\kappa(\kappa\sqrt{\hat{x}}-\bar{\mu})}{2\bar{\sigma}^2\sqrt{\hat{x}}}\right)^2\nonumber\\
=&2l(\hat{x}).
\end{align}
Therefore, we have
\begin{align}
\left.\frac{\partial^2 F(x)}{\partial x^2}\right|_{x=\hat{x}}=&\frac{B}{2}\left(\-\frac{1}{\ln2(1+\hat{x})^2}e^{-\frac{(\kappa\sqrt{\hat{x}}-\bar{\mu})^2}{2\bar{\sigma}^2}}-l(\hat{x})\right.\nonumber\\
&\left.-\log_2(1+\hat{x})e^{-\frac{(\kappa\sqrt{\hat{x}}-\bar{\mu})^2}{2\bar{\sigma}^2}}\frac{\kappa\bar{\mu}}{4\bar{\sigma}^2\hat{x}^{\frac{3}{2}}}\right).
\end{align}
Because $\kappa\sqrt{\hat{x}}-\bar{\mu}\geq0$, we can obtain that $ \left.\frac{\partial^2 F(x)}{\partial x^2}\right|_{x=\hat{x}}<0$. According to Lemma 2, $F(x)$ is a quasiconcave function. Furthermore, the solution to equation \eqref{Eq8} is unique.

\section{Proof of Lemma 3}
When $x=0$, we can obtain that
\begin{align}
\left.\frac{\partial \check{T}}{\partial x}\right|_{x=0}=&\frac{1}{\ln2}>0.
\end{align}
We also have $\check{T}=0$ when $x=0$, and $\check{T}>0$ when $x>0$. Because $\check{T}$ is a concave function with respect to $x$ for $x\in\left[0,\check{x}\right]$,
if $\left.\frac{\partial \check{T}}{\partial x}\right|_{x=\check{x}}\geq0$, we know $\check{T}$ is an increasing function with respect to $x$ for $x\in\left[0,\check{x}\right]$ and  $x$ that maximizes $\check{T}$ is  $\check{x}$. If $\left.\frac{\partial \check{T}}{\partial x}\right|_{x=\check{x}}<0$, we know that $\bar{T}$ is a function that first increases and then decreases for $x\in\left[0,\check{x}\right]$ and  $x$ that makes the first-order derivative  of $\check{T}$ with respect to $x$ to be 0.

\section{Proof of Lemma 4}
Letting $x=2^R-1(x\geq 0)$, the objective function $ \check{T}^{a}$  is
\begin{align}
H(x)=B\log_2(1+x)e^{-\Gamma x}.
\end{align}
Taking the first-order derivative of $H(x)$ with respect to $x$, we have
\begin{align}\label{Hq1}
\frac{\partial H(x)}{\partial x}=Be^{-\Gamma x}\left(\frac{1}{\ln2(1+x)}-\Gamma \log_2(1+x)\right).
\end{align}
From \eqref{Hq1}, we can readily know that $\frac{1}{\ln2(1+x)}$ is a monotonically decreasing function with respect to $x$, and $\log_2(1+x)$ is a monotonically increasing function with respect to $x$ when $x\geq 0$. We also have $\frac{1}{\ln2(1+x)}=\frac{1}{\ln2}$ and $\log_2(1+x)=0$ when $x=0$, and $\frac{1}{\ln2(1+x)}=0$ and $\log_2(1+x)=+\infty$ when $x=+\infty$. $\frac{\partial H(x)}{\partial x}$ is larger than 0 over the interval $x\in[0,x^\star)$, and smaller than 0 over the interval $x\in(x^\star, +\infty]$, where $x^\star$ satisfies    $\left.\frac{\partial F(x)}{\partial x}\right|_{x=x^\star}=0$, and the optimal $R$ is
 \begin{align}
\check{R}^\star=\frac{1}{\ln2}\mathcal{W}\left(1/\Gamma \right),
\end{align}
where $\mathcal{W}(z)$ represents the Lambert $\mathcal{W}$ function.

\end{document}